\documentclass[aps,prl,twocolumn,superscriptaddress,amsmath,amssymb]{revtex4-1}
\usepackage{graphicx}
\usepackage{dcolumn}
\usepackage{bm}
\usepackage{color}
\usepackage[T1]{fontenc}
\usepackage{mathptmx}

\begin{document}
\graphicspath{{fig/}}
\preprint{APS/123-QED}

\title{Topologically protected valley-dependent quantum photonic circuits}

\author{Yang Chen}
\thanks{These authors contributed equally to this work}
\affiliation{Key Laboratory of Quantum Information, University of Science and Technology of China, CAS, Hefei 230026, China}
\affiliation{Synergetic Innovation Center of Quantum Information \& Quantum Physics, University of Science and Technology of China, Hefei, Anhui 230026, China}
 
\author{Xin-Tao He}%
\thanks{These authors contributed equally to this work}
\affiliation{School of Physics and State Key Laboratory of Optoelectronic Materials and Technologies, Sun Yat-sen University, Guangzhou 510275, China}

\author{Yu-Jie Cheng}
\affiliation{Key Laboratory of Quantum Information, University of Science and Technology of China, CAS, Hefei 230026, China}
\affiliation{Synergetic Innovation Center of Quantum Information \& Quantum Physics, University of Science and Technology of China, Hefei, Anhui 230026, China}

\author{Hao-Yang Qiu}%
\affiliation{School of Physics and State Key Laboratory of Optoelectronic Materials and Technologies, Sun Yat-sen University, Guangzhou 510275, China}

\author{Lan-Tian Feng}
\affiliation{Key Laboratory of Quantum Information, University of Science and Technology of China, CAS, Hefei 230026, China}
\affiliation{Synergetic Innovation Center of Quantum Information \& Quantum Physics, University of Science and Technology of China, Hefei, Anhui 230026, China}

\author{Ming Zhang}
\affiliation{State Key Laboratory for Modern Optical Instrumentation, Centre for Optical and Electromagnetic Research, Zhejiang Provincial Key Laboratory for Sensing Technologies, Zhejiang University, Zijingang Campus, Hangzhou 310058, China}
\affiliation{Ningbo Research Institute, Zhejiang University, Ningbo 315100, China}

\author{Dao-Xin Dai}
\affiliation{State Key Laboratory for Modern Optical Instrumentation, Centre for Optical and Electromagnetic Research, Zhejiang Provincial Key Laboratory for Sensing Technologies, Zhejiang University, Zijingang Campus, Hangzhou 310058, China}
\affiliation{Ningbo Research Institute, Zhejiang University, Ningbo 315100, China}

\author{Guang-Can Guo}
\affiliation{Key Laboratory of Quantum Information, University of Science and Technology of China, CAS, Hefei 230026, China}
\affiliation{Synergetic Innovation Center of Quantum Information \& Quantum Physics, University of Science and Technology of China, Hefei, Anhui 230026, China}

\author{Jian-Wen Dong}%
\email{dongjwen@mail.sysu.edu.cn}
\affiliation{School of Physics and State Key Laboratory of Optoelectronic Materials and Technologies, Sun Yat-sen University, Guangzhou 510275, China}

\author{Xi-Feng Ren}%
\email{renxf@ustc.edu.cn}
\affiliation{Key Laboratory of Quantum Information, University of Science and Technology of China, CAS, Hefei 230026, China}
\affiliation{Synergetic Innovation Center of Quantum Information \& Quantum Physics, University of Science and Technology of China, Hefei, Anhui 230026, China}

%
%

\date{\today}

\begin{abstract}
Topological photonics has been introduced as a powerful platform for integrated optics, since it can deal with robust light transport, and be further extended to the quantum world. Strikingly, valley-contrasting physics in topological photonic structures contributes to valley-related edge states, their unidirectional coupling, and even valley-dependent wave-division in topological junctions. Here, we design and fabricate nanophotonic topological harpoon-shaped beam splitters (HSBSs) based on $120$-deg-bending interfaces and demonstrate the first on-chip valley-dependent quantum information process. Two-photon quantum interference, namely, Hong-Ou-Mandel (HOM) interference with a high visibility of $0.956\pm0.006$, is realized with our $50/50$ HSBS, which is constructed by two topologically distinct domain walls. Cascading this kind of HSBS together, we also demonstrate a simple quantum photonic circuit and generation of a path-entangled state. Our work shows that the photonic valley state can be used in quantum information processing, and it is possible to realize more complex quantum circuits with valley-dependent photonic topological insulators, which provides a novel method for on-chip quantum information processing.

\end{abstract}

\pacs{Valid PACS appear here}
\maketitle


Topological states of light provide an efficient way to encode information in silicon-on-insulator (SOI) slabs, particularly for the recent advances in topological light manipulation in photonic crystals (PCs). Research into two-dimensional photonic topological insulators (PTIs) in recent years has opened up intriguing areas from theoretical verification to technical applications, including robust edge state transport \cite{haldane2008possible, wang2009observation, rechtsman2013photonic, rechtsman2013topological}, optical delay lines \cite{hafezi2011robust}, topologically protected lasing effects \cite{Harari2018Topological, Bandres2018Topological}, and topological slow light \cite{guglielmon2019broadband, yang2013experimental}. Interestingly, the valley-dependent helical edge states travel in opposite directions with the corresponding circular polarizations, known as valley-Hall edge transport, which can be realized by breaking the spatial inversion symmetry of the system \cite{Mak2014The, Ju2015Topological, gorbachev2014detecting, xiao2007valley}.

The key part of a topological phase transition lies in opening an energy gap in the band structure at certain degenerate points by breaking either the time-reversal symmetry (TRS) or inversion symmetry \cite{bernevig2013topological}. PTIs without TRS have non-zero Chern numbers, which commonly requires an external or effective magnetic field for photons \cite{wang2009observation, Hafezi2013Imaging, Fang2012Realizing} or a temporal modulation of a photonic lattice \cite{rechtsman2013photonic}. On the other hand, in TRS systems, PTIs with specially tailored constructive parameters \cite{khanikaev2013photonic, chen2014experimental} and spatial configurations \cite{khanikaev2013photonic, chen2014experimental, he2019silicon, tian2020dispersion} can be readily accessible. By breaking the inversion symmetry, two-dimensional (2D) honeycomb lattice PCs with two inequivalent sublattices have been demonstrated to be a powerful platform to realize the latter, which can be related to the valley Hall effect with non-zero valley Chern numbers \cite{xiao2007valley, he2019silicon, yao2008valley}. Although systems with inversion symmetry breaking are time-reversal invariant, topological protection is manifested as long as disorder does not mix the valleys associated with the band \cite{he2019silicon, noh2018observation, Mikhail2018Robust}. 

In addition to the wide exploration of topological photonics towards classical waves, interesting physics could emerge by bringing topological photonics into the quantum world, including generation of quantum states \cite{Mittal2018ATS, blanco2018topological}, topologically protected unidirectional coupling of edge states by chiral quantum dots \cite{barik2018topological}, and topological protection of quantum coherence \cite{Wang:19, tambasco2018quantum, Nie2020}. More recently, on-chip Hong-Ou-Mandel (HOM) interference of topological boundary states with high visibility has been reported in a photonic waveguide array \cite{tambasco2018quantum}. Additionally, in a resonator array with coupled ring optical waveguides, the frequency-degenerate topological source of indistinguishable photon pairs has been tested by off-chip HOM interference with a beam splitter \cite{orre2020tunable}. However, the previous works usually used waveguide arrays to build topological photonic structures, which restricts the scaling up of circuits and convenient modulation of quantum states. More compact and scalable on-chip integrated quantum photonic operations with topologically-protected circuits remain to be established. Operating at a quarter-wavelength periodicity, a valley photonic crystal (VPC) waveguide provides a subwavelength strategy to explore topological photonic features. It is intriguing to apply the valley degree of freedom for on-chip quantum information processing with compact size, for which previous reports are lacking.

Here, we experimentally realize high visibility on-chip HOM interference at the junction of a valley-dependent harpoon-shaped topological interface. Two topologically-distinct domain walls arranged in a honeycomb lattice are used to form a ladder-like interface. Zigzag edges with a mid-gap energy \cite{noh2018observation} of the two domains make the "sides" of the ladder. Coupling linearly polarized light into the top and bottom domain walls of the valley-dependent photonic insulators, valley-dependent wave division is observed. Therefore, we obtain a $ 50/50 $ beam splitter shaped like a harpoon, named a harpoon-shaped beam splitter (HSBS). Based on the 120-deg-bending interfaces, we realize on-chip HOM interference in one HSBS with a high visibility of $ 0.956 $ and generation of the two-photon entangled state $ 1/\sqrt{2} \left(\left|20\right>-\left|02\right> \right) $ in valley-dependent quantum circuits by cascading two HSBSs. Compared to the previous works on the quantum interference in photonic waveguide arrays \cite{tambasco2018quantum}, our devices are CMOS compatible, scalable, and much more integrated, which guarantees the feasibility of extension to large-scale quantum information processing.

\begin{figure}[t]
\includegraphics[width=8.6cm]{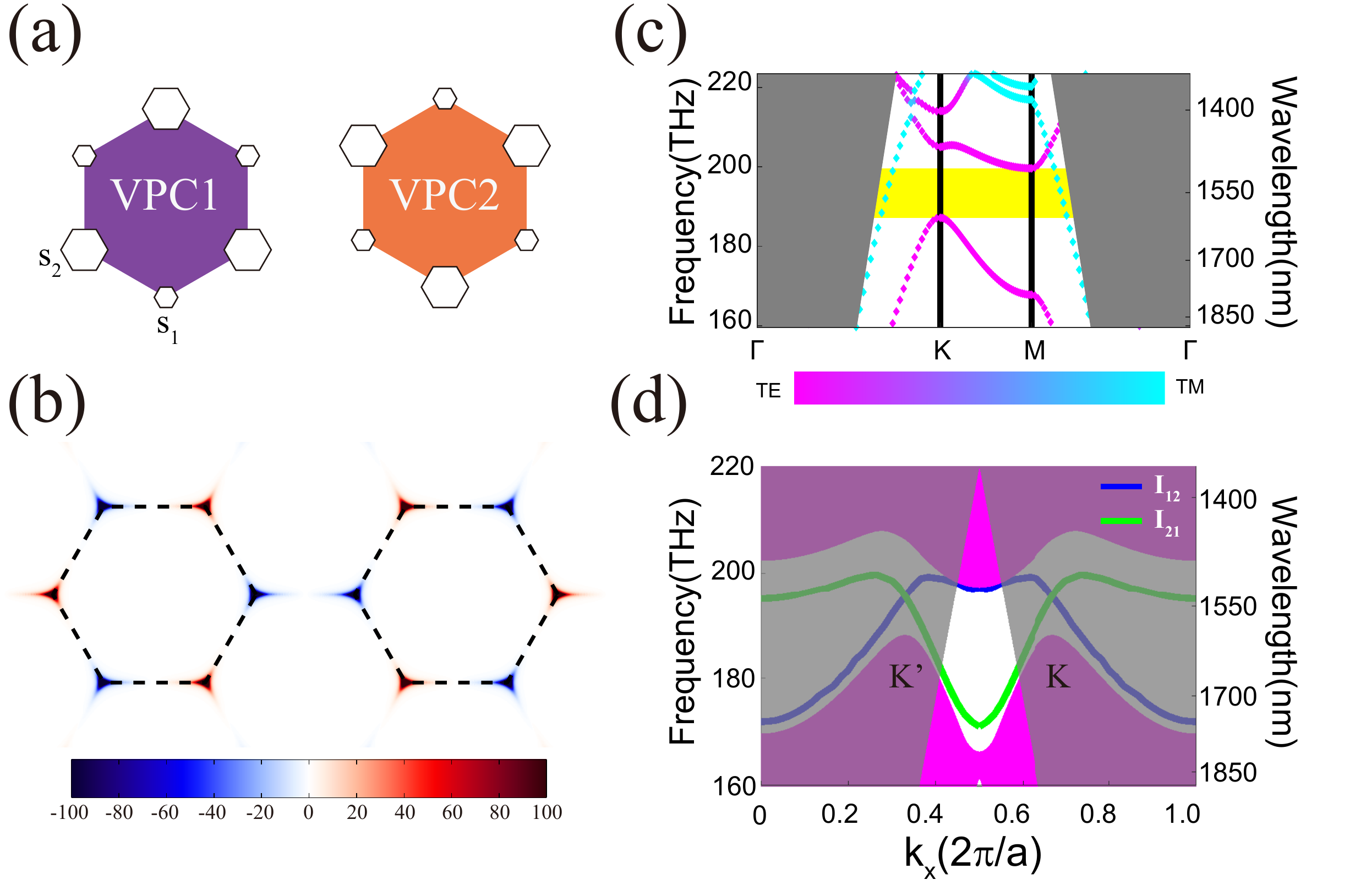}
\caption{\label{fig1} Topological valley photonic crystal and band structure. (a) Unit cell of valley photonic crystal $ 1 $ (VPC1) and VPC2. The sizes of the two hexagonal air holes are $ s_1 = 87 $ nm and $ s_2 = 127 $ nm. The lattice constant is $ a = 470 $ nm. (b) Distribution of TE1 Berry curvature for VPC1 (left) and VPC2 (right), which is localized near the corners of the first Brillouin zone. The peak of the Berry curvature is mainly localized around the $ K' (K) $ valley, while the sink is around the $ K (K') $ valley. (c) Bulk band for both VPC1 and VPC2, where the TE-like polarization bandgap lies between $ 1520 $ nm and $ 1600 $ nm. (d) Dispersion relations of topological interfaces $ I_{12} $ and $ I_{21} $ (shown in Fig. \ref{fig2}). The green and blue dotted line correspond to $ I_{12} $ and $ I_{21} $, respectively. For each edge state of the two topological interfaces, the states at the two non-equivalent valleys possess opposite-sign group velocities.}
\end{figure}

Our nanophotonic structures are fabricated on SOI wafers with $220$-nm-thick silicon layers by electron-beam lithography (for more details, see Supplementary Information). The valley-dependent photonic topological structures comprise two kinds of hexagonal-profile air holes of different side lengths arranged in a honeycomb lattice, which break the spatial-inversion symmetry of the system [shown in Fig. \ref{fig1} (a) and (b)]. First, we study the bulk topology of the TE-like band [shown in Fig. \ref{fig1} (c)] by using the MIT Photonic Bands (MPB) package \cite{johnson2001block} to calculate the band structures of the unit cell of the lattice [dotted line in Fig. \ref{fig1} (a) and (b)]. The side lengths of the two hexagonal air holes are $ s_1 = 87 $ nm and $ s_2 = 127 $ nm. The lattice constant is $ a = 470 $ nm. This special design opens a TE-like polarization bandgap between $ 1520 $ nm and $ 1600 $ nm.

\emph{The effective Hamiltonian.} ---In theory, our valley photonic crystal (VPC) can be approximatively described by an effective tight-binding Hamiltonian. Considering only the nearest-neighbour hopping, the tight-binding Hamiltonian is
\begin{equation}
H = -t \sum_{i \in A} \sum_{\bm{\delta}} \left( a_i^\dagger b_{i + \bm{\delta}} + b_{i + \bm{\delta}}^\dagger a_i \right) + \Delta \sum_i \left( a_i^\dagger a_i - b_i^\dagger b_i \right)
\end{equation}
where $a_i^\dagger \left( a_i \right)$ denotes the creation (annihilation) operator of photons on sublattice $A_i$. The first term describes the nearest-neighbour hopping, where the summation $ i $ runs over all the sublattices $ A_i $ and the sum over $ \bm{\delta} $ is carried out over the nearest-neighbour vectors. The second term denotes the energy difference $ 2\Delta $ between sublattices $ A $ and $ B $, which relates to the spatial-inversion symmetry breaking (see Supplementary Information for more details).

Diagonalization of the Hamiltonian and expansion of $ H_{\bm{k}} $ near the corners of the first Brillouin zone ($ K/K' $) reduces it to the two-dimensional Dirac equation $ H_{\bm{k}} = -\sqrt{3}/2 a t \left( q_x \tau_z \bm{\sigma}_x + q_y \bm{\sigma}_y \right) + \Delta \bm{\sigma}_z $ (see \cite{xiao2007valley} and Supplementary Information), where $ \bm{\sigma} $ is the Pauli matrix, and $ \bm{q} $ is the deviation from the Dirac points $ K $ and $ K' $ (denoted $ \tau_z = \pm 1 $). For such a 2D Dirac equation, the topological Chern number is given by $ C_{K/K'} = \tau_z \text{sgn} (\Delta)/2 $ \cite{shen2012topological}, which is defined near the Dirac points, and the valley Chern number is given by $ C_v =  C_K - C_{K'} = \text{sgn} (\Delta) $. To verify the valley Chern number in practical structures, an intuitive approach is to observe the simulated distribution of the Berry curvature. For VPC1 (VPC2), the peak of the Berry curvature is mainly localized around the $ K' (K) $ valley while the sink is around the $ K (K') $ valley [shown in Fig. \ref{fig1} (b)]. Therefore, the signs of the valley Chern number for VPC1 and VPC2 are opposite, which leads to valley-dependent edge states at the interfaces between the two topologically distinct domains.

\begin{figure}[t]
\includegraphics[width=8.6 cm]{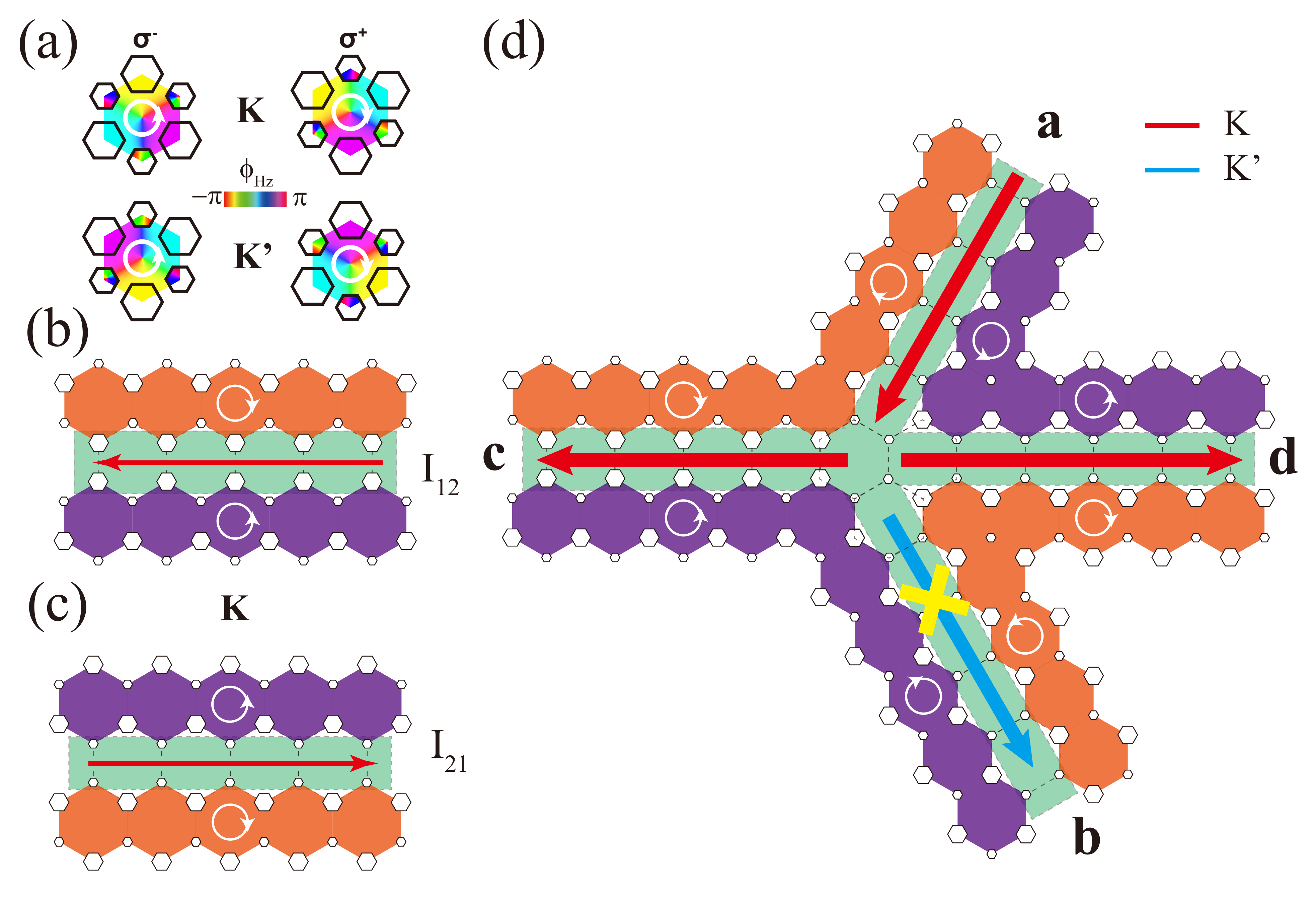}
\caption{\label{fig2} Phase vortex and selective coupling of valley-dependent edge states. (a) Electromagnetic phase vortex inside the unit cells of VPC1 and VPC2 at the TE1 band (simulated results). VPC1 supports clockwise (CW) and anti-clockwise (ACW) rotating states (phase increase direction) at the $ K' $ and $ K $ valleys, respectively, with the opposite results for VPC2. (b-c) Directional edge state transport of the two topologically distinct valley photonic crystals at the $ K' $ valley. The intensity vortex directions of VPC1 and VPC2 at the $ K $ valley are ACW and CW, respectively. This leads to valley-dependent backward (interface $ I_{12} $) or forward (interface $ I_{21} $) propagation of the edge states. The white circular arrows represent states at the TE1 band. More cases of directional edge state transport are available in the Supplementary Information. (d) Valley-dependent wave division at a topological junction. Photons are coupled into port $ a $ at the $ K $ valley (red arrow). Due to phase vortex matching, the propagating photons at the junction will couple into port $ c/d $ with the leftward (rightward) mode of $ I_{12} (I_{21}) $ at the $ K $ valley. However, the coupling to port $ b $ (blue arrow) is suppressed because the downward mode should be at the $ K' $ valley with the opposite phase vortex.}
\end{figure}

Locking of the valley state and the chirality of the phase vortex ensures selective coupling of edge states in topological valley photonic crystals (TVPCs) \cite{he2019silicon, tian2020dispersion, barik2018topological}. Here, the states at the two non-equivalent valleys $ K/K' $ play the role of spin, while the associated valley magnetic moment $ m\left( \bm{k} \right) $ determines the chirality of the phase vortex \cite{xiao2007valley}, with $ m(K,~K') = \tau_z \mu_B^* $ (see Supplementay Information), where $ \mu_B^* $ is the effective Bohr magneton at the bottom band [$ \text{sgn}(\mu_B^*) = \text{sgn}(\Delta) $; see Supplementary Information]. Thus, the unit cells shown in Fig. \ref{fig1} (a) and (b) at different valleys possess an intrinsic valley-dependent magnetic momentum. Note that these two configurations possess the same band structures [shown in Fig. \ref{fig1} (c)], but the motion of a photon in the two topologically non-trivial structures exhibits different physics at the two valleys. As has been extensively studied \cite{khanikaev2013photonic, he2019silicon, tian2020dispersion, yao2008valley, cheng2016robust, chen2017valley}, the orbital behaviour of photons in TVPCs is related to the flux intensity or the electromagnetic phase vortex inside the unit cell, which is symbolized as $ \sigma^{+/-} $ [the phase vortex of the TE1 band increases clockwise (CW) or anti-clockwise (ACW) by $ 2\pi $ as shown in Fig. \ref{fig2} (a) and Supplementary Information]. One feature of the valley-dependent edge states is that the orientation of the intensity vortex depends on the valley index and the configuration of the VPC, yielding valley-chirality locking of edge states [shown in Fig. \ref{fig2} (b) and (c) and Supplementary Information].

\begin{figure*}
\includegraphics[width=17.2 cm]{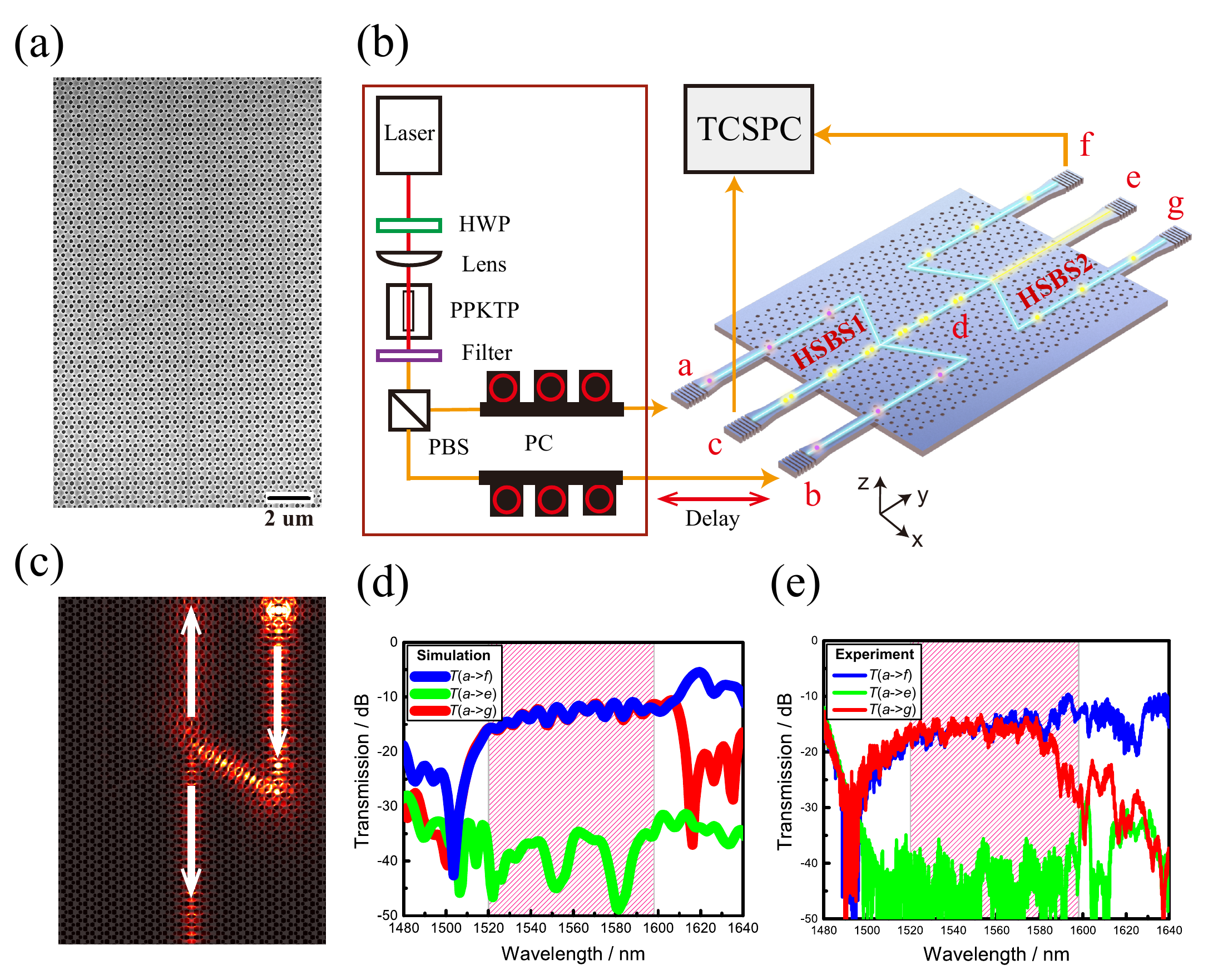}
\caption{\label{fig3} Topologically protected valley-dependent quantum circuits. (a) Scanning electron microscopy (SEM) image of the harpoon-shaped beam splitter (HSBS) constructed by four topologically distinct domains. (b) Experimental setup for the on-chip HOM interference. Photon pairs of $ 1550 $ nm are generated by pumping a non-linear PPKTP crystal. Here, the non-linear crystal is mounted on a thermoelectric cooler (TEC) board to fine tune the temperature. A long-pass filter is used to eliminate the laser signal. A polarization beam splitter (PBS) is used to separate the orthogonally polarized photons into two spatial modes. Polarization controllers (PCs) are used to adjust the polarization of the input photons. A time-correlated single photon counting (TCSPC) module is used to perform the coincidence measurements. (c) Simulation of the light evolution in our HSBS. As depicted by the white arrow, light is incident from the bottom side of the HSBS, and wave division is obtained. (d) Simulated and (e) measured transmission spectra for the quantum circuits shown in (b). Broadband $ 50/50 $ beam splitters in the bandgap range from $ 1520 $ nm to $ 1600 $ nm are obtained both theoretically and experimentally.}
\end{figure*}

\emph{Valley-dependent wave division.} ---Here, by stacking together the zigzag edges of VPC1 and VPC2, two types of a ladder-like interface are constructed, which are labelled $ I_{12} $ and $ I_{21} $ [shown in Fig. \ref{fig2} (b) and (c)]. The dispersion relations of the two interfaces are shown in Fig. \ref{fig1} (d). As depicted in Fig. \ref{fig2} (b) and (c), interface $ I_{12} $ ($I_{21}$) has negative (positive) velocity along the interface at the $ K $ valley, supporting the backward (forward)-propagating edge mode ($ I_{12}/I_{21} $), contrary to the cases at the $ K' $ valley (see Supplementary Information for more details). With the concept of valley-related directional transport along the ladder-like domain walls, it is simple to understand the coupling mechanism between multichannels, which is fundamental for generation of a two-photon entangled state in quantum optics. As shown in Fig. \ref{fig2} (d), we construct a four-channel structure based on the two types of interfaces. Here, the neighbouring domains possess distinct valley Chern numbers, thus resulting in selective coupling of valley-dependent edge states. For example, when photons are incident into port $ a $, they will couple to the downward edge mode at the $ K $ valley. Due to phase vortex matching, the propagating photons at the junction will couple into port $ c $ ($ d $) with the leftward (rightward) mode of $ I_{12} $ ($ I_{21} $) at the $ K $ valley. However, the coupling to port $ b $ is suppressed because the downward mode should be at the $ K' $ valley with the opposite phase vortex. Similarly, the results for incidence into port $ b (c/d) $ are also shown in the Supplementary Information. Therefore, valley-dependent wave division is formed as previously demonstrated \cite{chen2014experimental, tian2020dispersion}. We provide videos of FDTD simulations \cite{oskooi2010meep} of the process of wave-division in our system (see Supplementary Information). As shown in Fig. \ref{fig2} (d), when photons are incident into port $ a $, the structure of the topological interface between the transmitted (c, big air holes) and reflected (d, small air holes) arms are different, which is an asymmetrical beam splitting phenomenon. Thus the structure in Fig. \ref{fig2} (d) has two beam splitting phenomena: (i) symmetrical structure with a splitting ratio of 1:1 over a bandwidth of $80$ nm [Fig. \ref{fig3} (d) and (e)], (ii) asymmetrical structure with the splitting ratio is designed to be 1:1 at the wavelength of our photon source (see Supplementary Information).

Based on the discussion above, we construct harpoon-shaped beam splitters (HSBSs) by using sharp-bending interfaces, as shown in Fig. \ref{fig3} (a). We can further setup a more complex circuit by cascading two or more HSBSs together, as shown in Fig. \ref{fig3} (b). Here, with light being injected from port $ a $ or port $ b $ of HSBS1 [left part of the circuit in Fig. \ref{fig3} (b)], we measure the transmittance spectra of HSBS2 (right part of the circuit). As shown in Fig. \ref{fig3} (d) and (e), we can see that in the bandgap range from $ 1520 $ nm to $ 1600 $ nm, a high intensity ratio between the top wall (bottom wall) and the right wall ($ I_f / I_e $ or $ I_g / I_e $) is obtained both theoretically and experimentally, which arises from the valley-dependent wave division in the topological junction. From both the theoretically calculated and measured spectra, we see that the output intensity ratio between the top wall and the bottom wall ($ I_f / I_g $) is almost equal to $ 1 $ in the bandgap, which can be viewed as the reflectivity and transmittance of the HSBS (see Supplementary Information for the simulation results). Due to mirror symmetry along the $ x $ axis [Fig. \ref{fig3} (b)] of the zigzag edges, this balanced property is easily obtanied. For the bearded edges\cite{he2019silicon}, which lacks mirror symmetry along the $ x $ axis, the balanced property cannot be stricty guaranteed. This balanced property of our topological HSBS ensures high visibility quantum interference on-chip. 

\emph{On-chip HOM interference.} ---Two-photon quantum interference, known as Hong-Ou-Mandel (HOM) interference, is a purely quantum-mechanical feature of fourth-order interference \cite{PhysRevLett.59.2044, Ghosh1987ObservationON}. When two identical photons enter two ports of a $ 50/50 $ beam splitter (BS) separately, both photons are found together in one or the other output port of the BS, and the cases in which either both photons are reflected or both are transmitted cancel out due to destructive interference. Two-photon HOM interference has been widely accepted as a paradigm for testing photon indistinguishability \cite{santori2002indistinguishable}, generation of multi-photon states \cite{Wang2016}, and large-scale quantum computation and quantum simulation \cite{Shen2017Deep, Qiang2018}. Heretofore, HOM interference has been experimentally realized for electrons \cite{Bocquillon2013}, surface plasmons \cite{Fakonas2014Twoplasmon, Chen2018QPN}, phonons \cite{Toyoda2015}, atoms \cite{Lopes2015Atomic}, and photons \cite{PhysRevLett.59.2044}.

To perform on-chip HOM interference in the HSBSs, $ 1550 $ nm degenerate photon pairs are generated by pumping a periodically poled KTP (PPKTP) crystal with a $ 775 $ nm continuous wave laser via type-II spontaneous down conversion [Fig. \ref{fig3} (b)]. Here, the crystal temperature is properly tuned to ensure the wavelength degeneracy of photon pairs. The orthogonally polarized photon pairs are further separated into two spatial modes by a polarizing beam splitter (PBS) and collected by single-mode fibres (for more details, see Supplementary Information). The indistinguishability of the photon source is obtained from an HOM interference measurement, with a high raw visibility of $ 0.965 \pm 0.002 $, and the coherence length is $ 1.23 \pm 0.01 $ mm (see Supplementary Information for more details). The collected photons with linear polarization along the $ x $-direction [shown in Fig. \ref{fig3} (b)] are first coupled into the SOI waveguide (section size is $ 470 $ nm $ \times $ $ 220 $ nm) with grating couplers and then coupled into the valley-dependent topological interface (VDTI) due to the mode overlap between the silicon waveguide and the VDTI. The output photons are collected by the output waveguide and the grating couplers. Subsequently, photons are detected with the superconducting single-photon detectors and analyzed by the time-correlated single photon counting module. 

\begin{figure}[t]
\includegraphics[width=8.6cm]{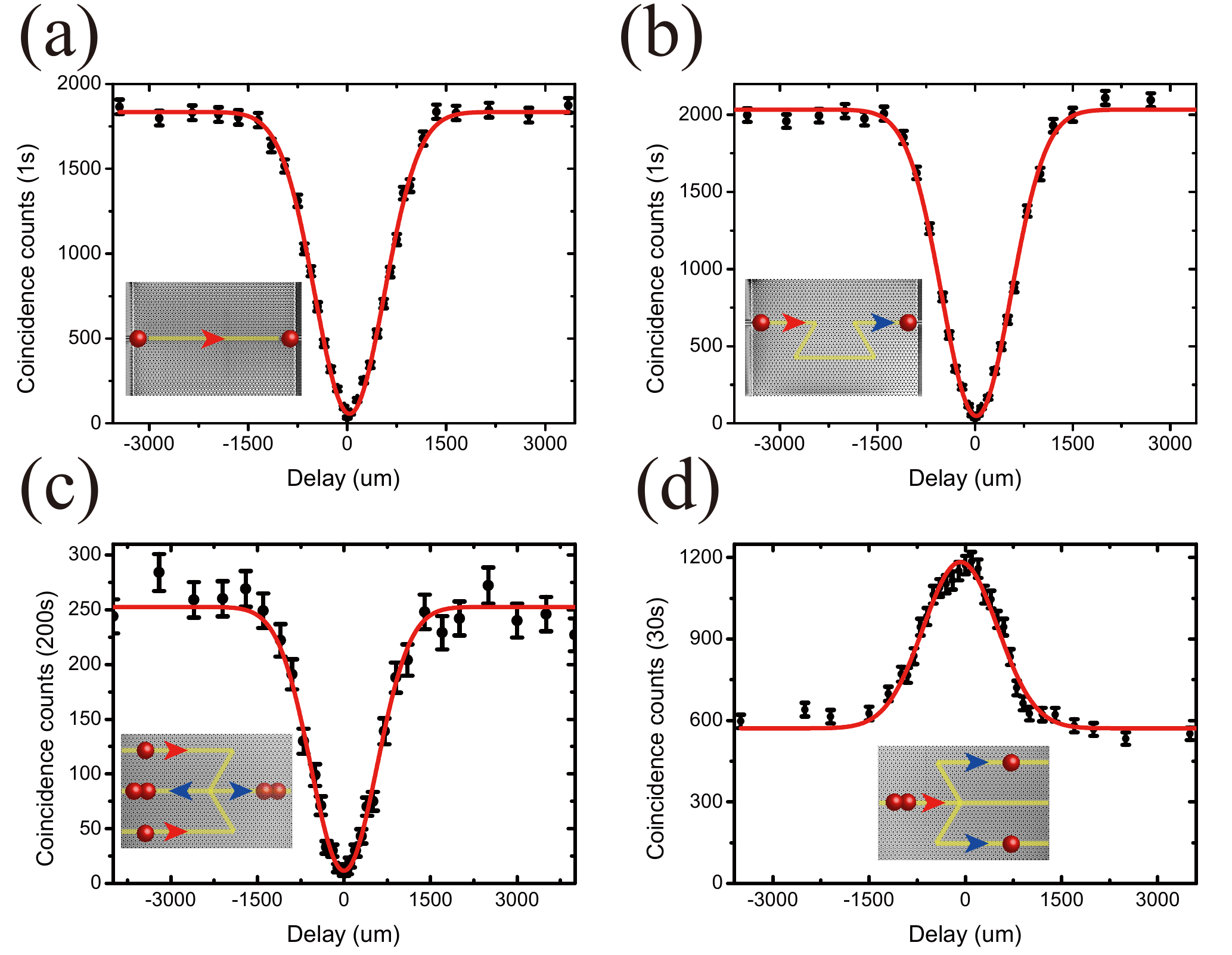}
\caption{\label{fig4} Measured HOM interference curve. Two-photon off-chip quantum interference using a fibre beam splitter for (a) the straight topological interface and (b) the $ \Omega $-shaped topological interface. The visibilities are $ 0.969 $ and $ 0.977 $, respectively. (c) On-chip HOM interference for the harpoon-shaped beam splitter (HSBS), with a visibility of $ 0.956 \pm 0.006 $. Coincidence measurements are performed between port $ c $ and port $ f $. (d) Coincidence measurements between port $ f $ and port $ g $ of the quantum circuit. We obtain an interference curve with a peak pattern. The interference visibility is $ 0.999 \pm 0.049 $. The error bars are all calculated by assuming Poisson statistics.}
\end{figure}

We first inject one photon into various configurations, including flat, $ Z $-shaped, and $ \Omega $-shaped topological interfaces \cite{he2019silicon}, and the other photon into a single mode fibre. The output photon pairs are further separately injected into two input ports of the $ 50/50 $ fibre beam splitter to perform the off-chip two photon quantum interference. We obtain high interference visibility for these various configurations, and all are above $ 0.90 $ (as depicted in Fig. \ref{fig4} and Supplementary Information), proving the indistinguishability of the photons transmitted through the topological interfaces with and without sharp turns.

Then, by injecting the down-converted photon pairs into the two arms of the HSBS (port $ a $ and port $ b $), we realize on-chip two photon quantum interference in the valley-dependent HSBS. As shown in Fig. \ref{fig4} (c), we obtain an HOM dip (coincidence measurements between port $ f $ and port $ c $) with a high raw visibility of $ 0.956 \pm 0.006 $, which is far beyond the classical interference limit of $ 0.5 $ \cite{Ghosh1987ObservationON}, and the coherence length is $ 1.29 \pm 0.04 $ mm (error bars are calculated by assuming Poisson statistics). This confirms that the two photons at the junction of HSBS1 after propagating along the topological interface are highly indistinguishable. Particularly, after the photon pairs interfere at the junction of HSB1, the path-entangled photon state $ 1/\sqrt{2} \left(\left|2_c, 0_d\right>-\left|0_c, 2_d\right> \right) $ will be generated. 

Furthermore, we show the scalability of the topological circuits. This can be confirmed by connecting output port $ d $ of HSBS1 with the input port of HSBS2. An HOM interference peak is expected to be observed by performing coincidence measurements between port $ f $ and port $ g $. The observed raw visibility is $ 0.999 \pm 0.049 $ [shown in Fig. \ref{fig4} (d)], indicating the generation of the two-photon state $ \left|2_c, 0_d\right> $. Considering the symmetry of paths $ c $ and $ d $, we can assume that the two-photon entangled state $ 1/\sqrt{2} \left(\left|20\right>-\left|02\right> \right) $ can be generated with the present circuit.

In summary, robust edge state transport of single photons along a topological interface with and without sharp turns is verified. We obtain a $ 50/50 $ topologically protected valley-dependent beam splitter constructed by $ 120 $-deg-bending interfaces between two topologically distinct domain walls, and we also experimentally realize on-chip quantum interference in these photonic valley-dependent topological insulators with high interference visibility. Finally, we further show the scalability of our structures in a circuit constructed by cascading two HSBSs. Our structure provides an accessible platform for quantum simulation of various topological phenomena in solid physics and will be beneficial for large-scale quantum information processing with more complex circuits \cite{Shen2017Deep, Qiang2018}.

\begin{acknowledgments}
This research is supported by the National Natural Science Foundation of China (NSFC) (Nos. 61590932, 62035016, 61775243, 11774333, 11904421, 62061160487), the Anhui Initiative in Quantum Information Technologies (No. AHY130300), the Strategic Priority Research Program of the Chinese Academy of Sciences (No. XDB24030601), the National Key R \& D Program (Nos. 2016YFA0301700, 2019YFB2203502), Natural Science Foundation of Guangdong Province (No. 2018B030308005), Guangzhou Science and Technology Program (No. 202002030322) and the Fundamental Research Funds for the Central Universities. This work is partially carried out at the USTC Center for Micro and Nanoscale Research and Fabrication.
\end{acknowledgments}

\nocite{*}

\bibliography{ref}

\providecommand{\noopsort}[1]{}\providecommand{\singleletter}[1]{#1}%
\begin{thebibliography}{47}%
\makeatletter
\providecommand \@ifxundefined [1]{%
 \@ifx{#1\undefined}
}%
\providecommand \@ifnum [1]{%
 \ifnum #1\expandafter \@firstoftwo
 \else \expandafter \@secondoftwo
 \fi
}%
\providecommand \@ifx [1]{%
 \ifx #1\expandafter \@firstoftwo
 \else \expandafter \@secondoftwo
 \fi
}%
\providecommand \natexlab [1]{#1}%
\providecommand \enquote  [1]{``#1''}%
\providecommand \bibnamefont  [1]{#1}%
\providecommand \bibfnamefont [1]{#1}%
\providecommand \citenamefont [1]{#1}%
\providecommand \href@noop [0]{\@secondoftwo}%
\providecommand \href [0]{\begingroup \@sanitize@url \@href}%
\providecommand \@href[1]{\@@startlink{#1}\@@href}%
\providecommand \@@href[1]{\endgroup#1\@@endlink}%
\providecommand \@sanitize@url [0]{\catcode `\\12\catcode `\$12\catcode
  `\&12\catcode `\#12\catcode `\^12\catcode `\_12\catcode `\%12\relax}%
\providecommand \@@startlink[1]{}%
\providecommand \@@endlink[0]{}%
\providecommand \url  [0]{\begingroup\@sanitize@url \@url }%
\providecommand \@url [1]{\endgroup\@href {#1}{\urlprefix }}%
\providecommand \urlprefix  [0]{URL }%
\providecommand \Eprint [0]{\href }%
\providecommand \doibase [0]{http://dx.doi.org/}%
\providecommand \selectlanguage [0]{\@gobble}%
\providecommand \bibinfo  [0]{\@secondoftwo}%
\providecommand \bibfield  [0]{\@secondoftwo}%
\providecommand \translation [1]{[#1]}%
\providecommand \BibitemOpen [0]{}%
\providecommand \bibitemStop [0]{}%
\providecommand \bibitemNoStop [0]{.\EOS\space}%
\providecommand \EOS [0]{\spacefactor3000\relax}%
\providecommand \BibitemShut  [1]{\csname bibitem#1\endcsname}%
\let\auto@bib@innerbib\@empty
\bibitem [{\citenamefont {Haldane}\ and\ \citenamefont
  {Raghu}(2008)}]{haldane2008possible}%
  \BibitemOpen
  \bibfield  {author} {\bibinfo {author} {\bibfnamefont {F.~D.~M.}\
  \bibnamefont {Haldane}}\ and\ \bibinfo {author} {\bibfnamefont
  {S.}~\bibnamefont {Raghu}},\ }\href@noop {} {\bibfield  {journal} {\bibinfo
  {journal} {Phys. Rev. Lett.}\ }\textbf {\bibinfo {volume} {100}},\ \bibinfo
  {pages} {013904} (\bibinfo {year} {2008})}\BibitemShut {NoStop}%
\bibitem [{\citenamefont {Wang}\ \emph {et~al.}(2009)\citenamefont {Wang},
  \citenamefont {Chong}, \citenamefont {Joannopoulos},\ and\ \citenamefont
  {Marin}}]{wang2009observation}%
  \BibitemOpen
  \bibfield  {author} {\bibinfo {author} {\bibfnamefont {Z.}~\bibnamefont
  {Wang}}, \bibinfo {author} {\bibfnamefont {Y.}~\bibnamefont {Chong}},
  \bibinfo {author} {\bibfnamefont {J.~D.}\ \bibnamefont {Joannopoulos}}, \
  and\ \bibinfo {author} {\bibfnamefont {S.}~\bibnamefont {Marin}},\
  }\href@noop {} {\bibfield  {journal} {\bibinfo  {journal} {Nature}\ }\textbf
  {\bibinfo {volume} {461}},\ \bibinfo {pages} {772} (\bibinfo {year}
  {2009})}\BibitemShut {NoStop}%
\bibitem [{\citenamefont {Rechtsman}\ \emph
  {et~al.}(2013{\natexlab{a}})\citenamefont {Rechtsman}, \citenamefont
  {Zeuner}, \citenamefont {Plotnik}, \citenamefont {Lumer}, \citenamefont
  {Podolsky}, \citenamefont {Dreisow}, \citenamefont {Nolte}, \citenamefont
  {Segev},\ and\ \citenamefont {Szameit}}]{rechtsman2013photonic}%
  \BibitemOpen
  \bibfield  {author} {\bibinfo {author} {\bibfnamefont {M.~C.}\ \bibnamefont
  {Rechtsman}}, \bibinfo {author} {\bibfnamefont {J.~M.}\ \bibnamefont
  {Zeuner}}, \bibinfo {author} {\bibfnamefont {Y.}~\bibnamefont {Plotnik}},
  \bibinfo {author} {\bibfnamefont {Y.}~\bibnamefont {Lumer}}, \bibinfo
  {author} {\bibfnamefont {D.}~\bibnamefont {Podolsky}}, \bibinfo {author}
  {\bibfnamefont {F.}~\bibnamefont {Dreisow}}, \bibinfo {author} {\bibfnamefont
  {S.}~\bibnamefont {Nolte}}, \bibinfo {author} {\bibfnamefont
  {M.}~\bibnamefont {Segev}}, \ and\ \bibinfo {author} {\bibfnamefont
  {A.}~\bibnamefont {Szameit}},\ }\href@noop {} {\bibfield  {journal} {\bibinfo
   {journal} {Nature}\ }\textbf {\bibinfo {volume} {496}},\ \bibinfo {pages}
  {196} (\bibinfo {year} {2013}{\natexlab{a}})}\BibitemShut {NoStop}%
\bibitem [{\citenamefont {Rechtsman}\ \emph
  {et~al.}(2013{\natexlab{b}})\citenamefont {Rechtsman}, \citenamefont
  {Plotnik}, \citenamefont {Zeuner}, \citenamefont {Song}, \citenamefont
  {Chen}, \citenamefont {Szameit},\ and\ \citenamefont
  {Segev}}]{rechtsman2013topological}%
  \BibitemOpen
  \bibfield  {author} {\bibinfo {author} {\bibfnamefont {M.~C.}\ \bibnamefont
  {Rechtsman}}, \bibinfo {author} {\bibfnamefont {Y.}~\bibnamefont {Plotnik}},
  \bibinfo {author} {\bibfnamefont {J.~M.}\ \bibnamefont {Zeuner}}, \bibinfo
  {author} {\bibfnamefont {D.}~\bibnamefont {Song}}, \bibinfo {author}
  {\bibfnamefont {Z.}~\bibnamefont {Chen}}, \bibinfo {author} {\bibfnamefont
  {A.}~\bibnamefont {Szameit}}, \ and\ \bibinfo {author} {\bibfnamefont
  {M.}~\bibnamefont {Segev}},\ }\href@noop {} {\bibfield  {journal} {\bibinfo
  {journal} {Phys. Rev. Lett.}\ }\textbf {\bibinfo {volume} {111}},\ \bibinfo
  {pages} {103901} (\bibinfo {year} {2013}{\natexlab{b}})}\BibitemShut
  {NoStop}%
\bibitem [{\citenamefont {Hafezi}\ \emph {et~al.}(2011)\citenamefont {Hafezi},
  \citenamefont {Demler}, \citenamefont {Lukin},\ and\ \citenamefont
  {Taylor}}]{hafezi2011robust}%
  \BibitemOpen
  \bibfield  {author} {\bibinfo {author} {\bibfnamefont {M.}~\bibnamefont
  {Hafezi}}, \bibinfo {author} {\bibfnamefont {E.~A.}\ \bibnamefont {Demler}},
  \bibinfo {author} {\bibfnamefont {M.~D.}\ \bibnamefont {Lukin}}, \ and\
  \bibinfo {author} {\bibfnamefont {J.~M.}\ \bibnamefont {Taylor}},\
  }\href@noop {} {\bibfield  {journal} {\bibinfo  {journal} {Nat. Phys.}\
  }\textbf {\bibinfo {volume} {7}},\ \bibinfo {pages} {907} (\bibinfo {year}
  {2011})}\BibitemShut {NoStop}%
\bibitem [{\citenamefont {Harari}\ \emph {et~al.}(2018)\citenamefont {Harari},
  \citenamefont {Bandres}, \citenamefont {Lumer}, \citenamefont {Rechtsman},
  \citenamefont {Chong}, \citenamefont {Khajavikhan}, \citenamefont
  {Christodoulides},\ and\ \citenamefont {Segev}}]{Harari2018Topological}%
  \BibitemOpen
  \bibfield  {author} {\bibinfo {author} {\bibfnamefont {G.}~\bibnamefont
  {Harari}}, \bibinfo {author} {\bibfnamefont {M.~A.}\ \bibnamefont {Bandres}},
  \bibinfo {author} {\bibfnamefont {Y.}~\bibnamefont {Lumer}}, \bibinfo
  {author} {\bibfnamefont {M.~C.}\ \bibnamefont {Rechtsman}}, \bibinfo {author}
  {\bibfnamefont {Y.~D.}\ \bibnamefont {Chong}}, \bibinfo {author}
  {\bibfnamefont {M.}~\bibnamefont {Khajavikhan}}, \bibinfo {author}
  {\bibfnamefont {D.~N.}\ \bibnamefont {Christodoulides}}, \ and\ \bibinfo
  {author} {\bibfnamefont {M.}~\bibnamefont {Segev}},\ }\href@noop {}
  {\bibfield  {journal} {\bibinfo  {journal} {Science}\ }\textbf {\bibinfo
  {volume} {359}},\ \bibinfo {pages} {6381} (\bibinfo {year}
  {2018})}\BibitemShut {NoStop}%
\bibitem [{\citenamefont {Bandres}\ \emph {et~al.}(2018)\citenamefont
  {Bandres}, \citenamefont {Wittek}, \citenamefont {Harari}, \citenamefont
  {Parto}, \citenamefont {Ren}, \citenamefont {Segev}, \citenamefont
  {Christodoulides},\ and\ \citenamefont
  {Khajavilchan}}]{Bandres2018Topological}%
  \BibitemOpen
  \bibfield  {author} {\bibinfo {author} {\bibfnamefont {M.~A.}\ \bibnamefont
  {Bandres}}, \bibinfo {author} {\bibfnamefont {S.}~\bibnamefont {Wittek}},
  \bibinfo {author} {\bibfnamefont {G.}~\bibnamefont {Harari}}, \bibinfo
  {author} {\bibfnamefont {M.}~\bibnamefont {Parto}}, \bibinfo {author}
  {\bibfnamefont {J.}~\bibnamefont {Ren}}, \bibinfo {author} {\bibfnamefont
  {M.}~\bibnamefont {Segev}}, \bibinfo {author} {\bibfnamefont {D.~N.}\
  \bibnamefont {Christodoulides}}, \ and\ \bibinfo {author} {\bibfnamefont
  {M.}~\bibnamefont {Khajavilchan}},\ }\href@noop {} {\bibfield  {journal}
  {\bibinfo  {journal} {Science}\ }\textbf {\bibinfo {volume} {359}},\ \bibinfo
  {pages} {1231} (\bibinfo {year} {2018})}\BibitemShut {NoStop}%
\bibitem [{\citenamefont {Guglielmon}\ and\ \citenamefont
  {Rechtsman}(2019)}]{guglielmon2019broadband}%
  \BibitemOpen
  \bibfield  {author} {\bibinfo {author} {\bibfnamefont {J.}~\bibnamefont
  {Guglielmon}}\ and\ \bibinfo {author} {\bibfnamefont {M.~C.}\ \bibnamefont
  {Rechtsman}},\ }\href@noop {} {\bibfield  {journal} {\bibinfo  {journal}
  {Phys. Rev. Lett.}\ }\textbf {\bibinfo {volume} {122}},\ \bibinfo {pages}
  {153904} (\bibinfo {year} {2019})}\BibitemShut {NoStop}%
\bibitem [{\citenamefont {Yang}\ \emph {et~al.}(2013)\citenamefont {Yang},
  \citenamefont {Poo}, \citenamefont {Wu}, \citenamefont {Gu},\ and\
  \citenamefont {Chen}}]{yang2013experimental}%
  \BibitemOpen
  \bibfield  {author} {\bibinfo {author} {\bibfnamefont {Y.}~\bibnamefont
  {Yang}}, \bibinfo {author} {\bibfnamefont {Y.}~\bibnamefont {Poo}}, \bibinfo
  {author} {\bibfnamefont {R.-x.}\ \bibnamefont {Wu}}, \bibinfo {author}
  {\bibfnamefont {Y.}~\bibnamefont {Gu}}, \ and\ \bibinfo {author}
  {\bibfnamefont {P.}~\bibnamefont {Chen}},\ }\href@noop {} {\bibfield
  {journal} {\bibinfo  {journal} {Appl. Phys. Lett.}\ }\textbf {\bibinfo
  {volume} {102}},\ \bibinfo {pages} {231113} (\bibinfo {year}
  {2013})}\BibitemShut {NoStop}%
\bibitem [{\citenamefont {Mak}\ \emph {et~al.}(2014)\citenamefont {Mak},
  \citenamefont {Mcgill}, \citenamefont {Park},\ and\ \citenamefont
  {Mceuen}}]{Mak2014The}%
  \BibitemOpen
  \bibfield  {author} {\bibinfo {author} {\bibfnamefont {K.~F.}\ \bibnamefont
  {Mak}}, \bibinfo {author} {\bibfnamefont {K.~L.}\ \bibnamefont {Mcgill}},
  \bibinfo {author} {\bibfnamefont {J.}~\bibnamefont {Park}}, \ and\ \bibinfo
  {author} {\bibfnamefont {P.~L.}\ \bibnamefont {Mceuen}},\ }\href@noop {}
  {\bibfield  {journal} {\bibinfo  {journal} {Science}\ }\textbf {\bibinfo
  {volume} {344}},\ \bibinfo {pages} {1489} (\bibinfo {year}
  {2014})}\BibitemShut {NoStop}%
\bibitem [{\citenamefont {Ju}\ \emph {et~al.}(2015)\citenamefont {Ju},
  \citenamefont {Shi}, \citenamefont {Nair}, \citenamefont {Lv}, \citenamefont
  {Jin}, \citenamefont {Velasco}, \citenamefont {Ojeda-Aristizabal},
  \citenamefont {Bechtel}, \citenamefont {Martin},\ and\ \citenamefont
  {Zettl}}]{Ju2015Topological}%
  \BibitemOpen
  \bibfield  {author} {\bibinfo {author} {\bibfnamefont {L.}~\bibnamefont
  {Ju}}, \bibinfo {author} {\bibfnamefont {Z.}~\bibnamefont {Shi}}, \bibinfo
  {author} {\bibfnamefont {N.}~\bibnamefont {Nair}}, \bibinfo {author}
  {\bibfnamefont {Y.}~\bibnamefont {Lv}}, \bibinfo {author} {\bibfnamefont
  {C.}~\bibnamefont {Jin}}, \bibinfo {author} {\bibfnamefont {J.}~\bibnamefont
  {Velasco}, \bibfnamefont {Jairo}}, \bibinfo {author} {\bibfnamefont
  {C.}~\bibnamefont {Ojeda-Aristizabal}}, \bibinfo {author} {\bibfnamefont
  {H.~A.}\ \bibnamefont {Bechtel}}, \bibinfo {author} {\bibfnamefont {M.~C.}\
  \bibnamefont {Martin}}, \ and\ \bibinfo {author} {\bibfnamefont
  {A.}~\bibnamefont {Zettl}},\ }\href@noop {} {\bibfield  {journal} {\bibinfo
  {journal} {Nature}\ }\textbf {\bibinfo {volume} {520}},\ \bibinfo {pages}
  {650} (\bibinfo {year} {2015})}\BibitemShut {NoStop}%
\bibitem [{\citenamefont {Gorbachev}\ \emph {et~al.}(2014)\citenamefont
  {Gorbachev}, \citenamefont {Song}, \citenamefont {Yu}, \citenamefont
  {Kretinin}, \citenamefont {Withers}, \citenamefont {Cao}, \citenamefont
  {Mishchenko}, \citenamefont {Grigorieva}, \citenamefont {Novoselov},
  \citenamefont {Levitov} \emph {et~al.}}]{gorbachev2014detecting}%
  \BibitemOpen
  \bibfield  {author} {\bibinfo {author} {\bibfnamefont {R.~V.}\ \bibnamefont
  {Gorbachev}}, \bibinfo {author} {\bibfnamefont {J.~C.~W.}\ \bibnamefont
  {Song}}, \bibinfo {author} {\bibfnamefont {G.}~\bibnamefont {Yu}}, \bibinfo
  {author} {\bibfnamefont {A.~V.}\ \bibnamefont {Kretinin}}, \bibinfo {author}
  {\bibfnamefont {F.}~\bibnamefont {Withers}}, \bibinfo {author} {\bibfnamefont
  {Y.}~\bibnamefont {Cao}}, \bibinfo {author} {\bibfnamefont {A.}~\bibnamefont
  {Mishchenko}}, \bibinfo {author} {\bibfnamefont {I.~V.}\ \bibnamefont
  {Grigorieva}}, \bibinfo {author} {\bibfnamefont {K.~S.}\ \bibnamefont
  {Novoselov}}, \bibinfo {author} {\bibfnamefont {L.~S.}\ \bibnamefont
  {Levitov}},  \emph {et~al.},\ }\href@noop {} {\bibfield  {journal} {\bibinfo
  {journal} {Science}\ }\textbf {\bibinfo {volume} {346}},\ \bibinfo {pages}
  {448} (\bibinfo {year} {2014})}\BibitemShut {NoStop}%
\bibitem [{\citenamefont {Xiao}\ \emph {et~al.}(2007)\citenamefont {Xiao},
  \citenamefont {Yao},\ and\ \citenamefont {Niu}}]{xiao2007valley}%
  \BibitemOpen
  \bibfield  {author} {\bibinfo {author} {\bibfnamefont {D.}~\bibnamefont
  {Xiao}}, \bibinfo {author} {\bibfnamefont {W.}~\bibnamefont {Yao}}, \ and\
  \bibinfo {author} {\bibfnamefont {Q.}~\bibnamefont {Niu}},\ }\href@noop {}
  {\bibfield  {journal} {\bibinfo  {journal} {Phys. Rev. Lett.}\ }\textbf
  {\bibinfo {volume} {99}},\ \bibinfo {pages} {236809} (\bibinfo {year}
  {2007})}\BibitemShut {NoStop}%
\bibitem [{\citenamefont {Bernevig}\ and\ \citenamefont
  {Hughes}(2013)}]{bernevig2013topological}%
  \BibitemOpen
  \bibfield  {author} {\bibinfo {author} {\bibfnamefont {B.~A.}\ \bibnamefont
  {Bernevig}}\ and\ \bibinfo {author} {\bibfnamefont {T.~L.}\ \bibnamefont
  {Hughes}},\ }\href@noop {} {\emph {\bibinfo {title} {Topological insulators
  and topological superconductors}}}\ (\bibinfo  {publisher} {Princeton
  university press},\ \bibinfo {year} {2013})\BibitemShut {NoStop}%
\bibitem [{\citenamefont {Hafezi}\ \emph {et~al.}(2013)\citenamefont {Hafezi},
  \citenamefont {M.}, \citenamefont {Mittal}, \citenamefont {S.}, \citenamefont
  {Fan}, \citenamefont {J.}, \citenamefont {Migdall}, \citenamefont {A.},
  \citenamefont {Taylor},\ and\ \citenamefont {and}}]{Hafezi2013Imaging}%
  \BibitemOpen
  \bibfield  {author} {\bibinfo {author} {\bibnamefont {Hafezi}}, \bibinfo
  {author} {\bibnamefont {M.}}, \bibinfo {author} {\bibnamefont {Mittal}},
  \bibinfo {author} {\bibnamefont {S.}}, \bibinfo {author} {\bibnamefont
  {Fan}}, \bibinfo {author} {\bibnamefont {J.}}, \bibinfo {author}
  {\bibnamefont {Migdall}}, \bibinfo {author} {\bibnamefont {A.}}, \bibinfo
  {author} {\bibnamefont {Taylor}}, \ and\ \bibinfo {author} {\bibfnamefont
  {J.}~\bibnamefont {and}},\ }\href@noop {} {\bibfield  {journal} {\bibinfo
  {journal} {Nat. Photonics}\ }\textbf {\bibinfo {volume} {7}},\ \bibinfo
  {pages} {1001} (\bibinfo {year} {2013})}\BibitemShut {NoStop}%
\bibitem [{\citenamefont {Fang}\ \emph {et~al.}(2012)\citenamefont {Fang},
  \citenamefont {Yu},\ and\ \citenamefont {Fan}}]{Fang2012Realizing}%
  \BibitemOpen
  \bibfield  {author} {\bibinfo {author} {\bibfnamefont {K.}~\bibnamefont
  {Fang}}, \bibinfo {author} {\bibfnamefont {Z.}~\bibnamefont {Yu}}, \ and\
  \bibinfo {author} {\bibfnamefont {S.}~\bibnamefont {Fan}},\ }\href@noop {}
  {\bibfield  {journal} {\bibinfo  {journal} {Nat. Photonics}\ }\textbf
  {\bibinfo {volume} {6}},\ \bibinfo {pages} {782} (\bibinfo {year}
  {2012})}\BibitemShut {NoStop}%
\bibitem [{\citenamefont {Khanikaev}\ \emph {et~al.}(2013)\citenamefont
  {Khanikaev}, \citenamefont {Mousavi}, \citenamefont {Tse}, \citenamefont
  {Kargarian}, \citenamefont {MacDonald},\ and\ \citenamefont
  {Shvets}}]{khanikaev2013photonic}%
  \BibitemOpen
  \bibfield  {author} {\bibinfo {author} {\bibfnamefont {A.~B.}\ \bibnamefont
  {Khanikaev}}, \bibinfo {author} {\bibfnamefont {S.~H.}\ \bibnamefont
  {Mousavi}}, \bibinfo {author} {\bibfnamefont {W.-K.}\ \bibnamefont {Tse}},
  \bibinfo {author} {\bibfnamefont {M.}~\bibnamefont {Kargarian}}, \bibinfo
  {author} {\bibfnamefont {A.~H.}\ \bibnamefont {MacDonald}}, \ and\ \bibinfo
  {author} {\bibfnamefont {G.}~\bibnamefont {Shvets}},\ }\href@noop {}
  {\bibfield  {journal} {\bibinfo  {journal} {Nat. Mater.}\ }\textbf {\bibinfo
  {volume} {12}},\ \bibinfo {pages} {233} (\bibinfo {year} {2013})}\BibitemShut
  {NoStop}%
\bibitem [{\citenamefont {Chen}\ \emph {et~al.}(2014)\citenamefont {Chen},
  \citenamefont {Jiang}, \citenamefont {Chen}, \citenamefont {Zhu},
  \citenamefont {Zhou}, \citenamefont {Dong},\ and\ \citenamefont
  {Chan}}]{chen2014experimental}%
  \BibitemOpen
  \bibfield  {author} {\bibinfo {author} {\bibfnamefont {W.-J.}\ \bibnamefont
  {Chen}}, \bibinfo {author} {\bibfnamefont {S.-J.}\ \bibnamefont {Jiang}},
  \bibinfo {author} {\bibfnamefont {X.-D.}\ \bibnamefont {Chen}}, \bibinfo
  {author} {\bibfnamefont {B.}~\bibnamefont {Zhu}}, \bibinfo {author}
  {\bibfnamefont {L.}~\bibnamefont {Zhou}}, \bibinfo {author} {\bibfnamefont
  {J.-W.}\ \bibnamefont {Dong}}, \ and\ \bibinfo {author} {\bibfnamefont
  {C.~T.}\ \bibnamefont {Chan}},\ }\href@noop {} {\bibfield  {journal}
  {\bibinfo  {journal} {Nat. Commun.}\ }\textbf {\bibinfo {volume} {5}},\
  \bibinfo {pages} {1} (\bibinfo {year} {2014})}\BibitemShut {NoStop}%
\bibitem [{\citenamefont {He}\ \emph {et~al.}(2019)\citenamefont {He},
  \citenamefont {Liang}, \citenamefont {Yuan}, \citenamefont {Qiu},
  \citenamefont {Chen}, \citenamefont {Zhao},\ and\ \citenamefont
  {Dong}}]{he2019silicon}%
  \BibitemOpen
  \bibfield  {author} {\bibinfo {author} {\bibfnamefont {X.-T.}\ \bibnamefont
  {He}}, \bibinfo {author} {\bibfnamefont {E.-T.}\ \bibnamefont {Liang}},
  \bibinfo {author} {\bibfnamefont {J.-J.}\ \bibnamefont {Yuan}}, \bibinfo
  {author} {\bibfnamefont {H.-Y.}\ \bibnamefont {Qiu}}, \bibinfo {author}
  {\bibfnamefont {X.-D.}\ \bibnamefont {Chen}}, \bibinfo {author}
  {\bibfnamefont {F.-L.}\ \bibnamefont {Zhao}}, \ and\ \bibinfo {author}
  {\bibfnamefont {J.-W.}\ \bibnamefont {Dong}},\ }\href@noop {} {\bibfield
  {journal} {\bibinfo  {journal} {Nat. Commun.}\ }\textbf {\bibinfo {volume}
  {10}},\ \bibinfo {pages} {1} (\bibinfo {year} {2019})}\BibitemShut {NoStop}%
\bibitem [{\citenamefont {Tian}\ \emph {et~al.}(2020)\citenamefont {Tian},
  \citenamefont {Shen}, \citenamefont {Li}, \citenamefont {Reit}, \citenamefont
  {Bachman}, \citenamefont {Socolar}, \citenamefont {Cummer},\ and\
  \citenamefont {Huang}}]{tian2020dispersion}%
  \BibitemOpen
  \bibfield  {author} {\bibinfo {author} {\bibfnamefont {Z.}~\bibnamefont
  {Tian}}, \bibinfo {author} {\bibfnamefont {C.}~\bibnamefont {Shen}}, \bibinfo
  {author} {\bibfnamefont {J.}~\bibnamefont {Li}}, \bibinfo {author}
  {\bibfnamefont {E.}~\bibnamefont {Reit}}, \bibinfo {author} {\bibfnamefont
  {H.}~\bibnamefont {Bachman}}, \bibinfo {author} {\bibfnamefont {J.~E.}\
  \bibnamefont {Socolar}}, \bibinfo {author} {\bibfnamefont {S.~A.}\
  \bibnamefont {Cummer}}, \ and\ \bibinfo {author} {\bibfnamefont {T.~J.}\
  \bibnamefont {Huang}},\ }\href@noop {} {\bibfield  {journal} {\bibinfo
  {journal} {Nat. Commun.}\ }\textbf {\bibinfo {volume} {11}},\ \bibinfo
  {pages} {1} (\bibinfo {year} {2020})}\BibitemShut {NoStop}%
\bibitem [{\citenamefont {Yao}\ \emph {et~al.}(2008)\citenamefont {Yao},
  \citenamefont {Xiao},\ and\ \citenamefont {Niu}}]{yao2008valley}%
  \BibitemOpen
  \bibfield  {author} {\bibinfo {author} {\bibfnamefont {W.}~\bibnamefont
  {Yao}}, \bibinfo {author} {\bibfnamefont {D.}~\bibnamefont {Xiao}}, \ and\
  \bibinfo {author} {\bibfnamefont {Q.}~\bibnamefont {Niu}},\ }\href@noop {}
  {\bibfield  {journal} {\bibinfo  {journal} {Phys. Rev. B}\ }\textbf {\bibinfo
  {volume} {77}},\ \bibinfo {pages} {235406} (\bibinfo {year}
  {2008})}\BibitemShut {NoStop}%
\bibitem [{\citenamefont {Noh}\ \emph {et~al.}(2018)\citenamefont {Noh},
  \citenamefont {Huang}, \citenamefont {Chen},\ and\ \citenamefont
  {Rechtsman}}]{noh2018observation}%
  \BibitemOpen
  \bibfield  {author} {\bibinfo {author} {\bibfnamefont {J.}~\bibnamefont
  {Noh}}, \bibinfo {author} {\bibfnamefont {S.}~\bibnamefont {Huang}}, \bibinfo
  {author} {\bibfnamefont {K.~P.}\ \bibnamefont {Chen}}, \ and\ \bibinfo
  {author} {\bibfnamefont {M.~C.}\ \bibnamefont {Rechtsman}},\ }\href@noop {}
  {\bibfield  {journal} {\bibinfo  {journal} {Phys. Rev. Lett.}\ }\textbf
  {\bibinfo {volume} {120}},\ \bibinfo {pages} {063902} (\bibinfo {year}
  {2018})}\BibitemShut {NoStop}%
\bibitem [{\citenamefont {Mikhail}\ \emph {et~al.}(2018)\citenamefont
  {Mikhail}, \citenamefont {I}, \citenamefont {Shalaev}, \citenamefont
  {Wiktor}, \citenamefont {Walasik}, \citenamefont {Alexander}, \citenamefont
  {Tsukernik}, \citenamefont {Yun}, \citenamefont {Xu},\ and\ \citenamefont
  {and}}]{Mikhail2018Robust}%
  \BibitemOpen
  \bibfield  {author} {\bibinfo {author} {\bibnamefont {Mikhail}}, \bibinfo
  {author} {\bibnamefont {I}}, \bibinfo {author} {\bibnamefont {Shalaev}},
  \bibinfo {author} {\bibnamefont {Wiktor}}, \bibinfo {author} {\bibnamefont
  {Walasik}}, \bibinfo {author} {\bibnamefont {Alexander}}, \bibinfo {author}
  {\bibnamefont {Tsukernik}}, \bibinfo {author} {\bibnamefont {Yun}}, \bibinfo
  {author} {\bibnamefont {Xu}}, \ and\ \bibinfo {author} {\bibfnamefont
  {N.}~\bibnamefont {and}},\ }\href@noop {} {\bibfield  {journal} {\bibinfo
  {journal} {Nat. Nanotech}\ }\textbf {\bibinfo {volume} {14}},\ \bibinfo
  {pages} {31} (\bibinfo {year} {2018})}\BibitemShut {NoStop}%
\bibitem [{\citenamefont {Mittal}\ \emph {et~al.}(2018)\citenamefont {Mittal},
  \citenamefont {Goldschmidt},\ and\ \citenamefont {Hafezi}}]{Mittal2018ATS}%
  \BibitemOpen
  \bibfield  {author} {\bibinfo {author} {\bibfnamefont {S.}~\bibnamefont
  {Mittal}}, \bibinfo {author} {\bibfnamefont {E.~A.}\ \bibnamefont
  {Goldschmidt}}, \ and\ \bibinfo {author} {\bibfnamefont {M.}~\bibnamefont
  {Hafezi}},\ }\href@noop {} {\bibfield  {journal} {\bibinfo  {journal}
  {Nature}\ }\textbf {\bibinfo {volume} {561}},\ \bibinfo {pages} {502}
  (\bibinfo {year} {2018})}\BibitemShut {NoStop}%
\bibitem [{\citenamefont {Blanco-Redondo}\ \emph {et~al.}(2018)\citenamefont
  {Blanco-Redondo}, \citenamefont {Bell}, \citenamefont {Oren}, \citenamefont
  {Eggleton},\ and\ \citenamefont {Segev}}]{blanco2018topological}%
  \BibitemOpen
  \bibfield  {author} {\bibinfo {author} {\bibfnamefont {A.}~\bibnamefont
  {Blanco-Redondo}}, \bibinfo {author} {\bibfnamefont {B.}~\bibnamefont
  {Bell}}, \bibinfo {author} {\bibfnamefont {D.}~\bibnamefont {Oren}}, \bibinfo
  {author} {\bibfnamefont {B.~J.}\ \bibnamefont {Eggleton}}, \ and\ \bibinfo
  {author} {\bibfnamefont {M.}~\bibnamefont {Segev}},\ }\href@noop {}
  {\bibfield  {journal} {\bibinfo  {journal} {Science}\ }\textbf {\bibinfo
  {volume} {362}},\ \bibinfo {pages} {568} (\bibinfo {year}
  {2018})}\BibitemShut {NoStop}%
\bibitem [{\citenamefont {Barik}\ \emph {et~al.}(2018)\citenamefont {Barik},
  \citenamefont {Karasahin}, \citenamefont {Flower}, \citenamefont {Cai},
  \citenamefont {Miyake}, \citenamefont {DeGottardi}, \citenamefont {Hafezi},\
  and\ \citenamefont {Waks}}]{barik2018topological}%
  \BibitemOpen
  \bibfield  {author} {\bibinfo {author} {\bibfnamefont {S.}~\bibnamefont
  {Barik}}, \bibinfo {author} {\bibfnamefont {A.}~\bibnamefont {Karasahin}},
  \bibinfo {author} {\bibfnamefont {C.}~\bibnamefont {Flower}}, \bibinfo
  {author} {\bibfnamefont {T.}~\bibnamefont {Cai}}, \bibinfo {author}
  {\bibfnamefont {H.}~\bibnamefont {Miyake}}, \bibinfo {author} {\bibfnamefont
  {W.}~\bibnamefont {DeGottardi}}, \bibinfo {author} {\bibfnamefont
  {M.}~\bibnamefont {Hafezi}}, \ and\ \bibinfo {author} {\bibfnamefont
  {E.}~\bibnamefont {Waks}},\ }\href@noop {} {\bibfield  {journal} {\bibinfo
  {journal} {Science}\ }\textbf {\bibinfo {volume} {359}},\ \bibinfo {pages}
  {666} (\bibinfo {year} {2018})}\BibitemShut {NoStop}%
\bibitem [{\citenamefont {Wang}\ \emph {et~al.}(2019)\citenamefont {Wang},
  \citenamefont {Pang}, \citenamefont {Lu}, \citenamefont {Gao}, \citenamefont
  {Chang}, \citenamefont {Qiao}, \citenamefont {Jiao}, \citenamefont {Tang},\
  and\ \citenamefont {Jin}}]{Wang:19}%
  \BibitemOpen
  \bibfield  {author} {\bibinfo {author} {\bibfnamefont {Y.}~\bibnamefont
  {Wang}}, \bibinfo {author} {\bibfnamefont {X.-L.}\ \bibnamefont {Pang}},
  \bibinfo {author} {\bibfnamefont {Y.-H.}\ \bibnamefont {Lu}}, \bibinfo
  {author} {\bibfnamefont {J.}~\bibnamefont {Gao}}, \bibinfo {author}
  {\bibfnamefont {Y.-J.}\ \bibnamefont {Chang}}, \bibinfo {author}
  {\bibfnamefont {L.-F.}\ \bibnamefont {Qiao}}, \bibinfo {author}
  {\bibfnamefont {Z.-Q.}\ \bibnamefont {Jiao}}, \bibinfo {author}
  {\bibfnamefont {H.}~\bibnamefont {Tang}}, \ and\ \bibinfo {author}
  {\bibfnamefont {X.-M.}\ \bibnamefont {Jin}},\ }\href {\doibase
  10.1364/OPTICA.6.000955} {\bibfield  {journal} {\bibinfo  {journal} {Optica}\
  }\textbf {\bibinfo {volume} {6}},\ \bibinfo {pages} {955} (\bibinfo {year}
  {2019})}\BibitemShut {NoStop}%
\bibitem [{\citenamefont {Tambasco}\ \emph {et~al.}(2018)\citenamefont
  {Tambasco}, \citenamefont {Corrielli}, \citenamefont {Chapman}, \citenamefont
  {Crespi}, \citenamefont {Zilberberg}, \citenamefont {Osellame},\ and\
  \citenamefont {Peruzzo}}]{tambasco2018quantum}%
  \BibitemOpen
  \bibfield  {author} {\bibinfo {author} {\bibfnamefont {J.-L.}\ \bibnamefont
  {Tambasco}}, \bibinfo {author} {\bibfnamefont {G.}~\bibnamefont {Corrielli}},
  \bibinfo {author} {\bibfnamefont {R.~J.}\ \bibnamefont {Chapman}}, \bibinfo
  {author} {\bibfnamefont {A.}~\bibnamefont {Crespi}}, \bibinfo {author}
  {\bibfnamefont {O.}~\bibnamefont {Zilberberg}}, \bibinfo {author}
  {\bibfnamefont {R.}~\bibnamefont {Osellame}}, \ and\ \bibinfo {author}
  {\bibfnamefont {A.}~\bibnamefont {Peruzzo}},\ }\href@noop {} {\bibfield
  {journal} {\bibinfo  {journal} {Sci. Adv.}\ }\textbf {\bibinfo {volume}
  {4}},\ \bibinfo {pages} {eaat3187} (\bibinfo {year} {2018})}\BibitemShut
  {NoStop}%
\bibitem [{\citenamefont {Nie}\ \emph {et~al.}(2020)\citenamefont {Nie},
  \citenamefont {Peng}, \citenamefont {Nori},\ and\ \citenamefont
  {xi~Liu}}]{Nie2020}%
  \BibitemOpen
  \bibfield  {author} {\bibinfo {author} {\bibfnamefont {W.}~\bibnamefont
  {Nie}}, \bibinfo {author} {\bibfnamefont {Z.~H.}\ \bibnamefont {Peng}},
  \bibinfo {author} {\bibfnamefont {F.}~\bibnamefont {Nori}}, \ and\ \bibinfo
  {author} {\bibfnamefont {Y.}~\bibnamefont {xi~Liu}},\ }\href@noop {}
  {\bibfield  {journal} {\bibinfo  {journal} {Phys. Rev. Lett.}\ }\textbf
  {\bibinfo {volume} {124 2}},\ \bibinfo {pages} {023603} (\bibinfo {year}
  {2020})}\BibitemShut {NoStop}%
\bibitem [{\citenamefont {Orre}\ \emph {et~al.}(2020)\citenamefont {Orre},
  \citenamefont {Mittal}, \citenamefont {Goldschmidt},\ and\ \citenamefont
  {Hafezi}}]{orre2020tunable}%
  \BibitemOpen
  \bibfield  {author} {\bibinfo {author} {\bibfnamefont {V.~V.}\ \bibnamefont
  {Orre}}, \bibinfo {author} {\bibfnamefont {S.}~\bibnamefont {Mittal}},
  \bibinfo {author} {\bibfnamefont {E.~A.}\ \bibnamefont {Goldschmidt}}, \ and\
  \bibinfo {author} {\bibfnamefont {M.}~\bibnamefont {Hafezi}},\ }\href@noop {}
  {\enquote {\bibinfo {title} {Tunable quantum interference using a topological
  source of indistinguishable photon pairs},}\ } (\bibinfo {year} {2020}),\
  \Eprint {http://arxiv.org/abs/2006.03084} {arXiv:2006.03084 [physics.optics]}
  \BibitemShut {NoStop}%
\bibitem [{\citenamefont {Johnson}\ and\ \citenamefont
  {Joannopoulos}(2001)}]{johnson2001block}%
  \BibitemOpen
  \bibfield  {author} {\bibinfo {author} {\bibfnamefont {S.~G.}\ \bibnamefont
  {Johnson}}\ and\ \bibinfo {author} {\bibfnamefont {J.~D.}\ \bibnamefont
  {Joannopoulos}},\ }\href@noop {} {\bibfield  {journal} {\bibinfo  {journal}
  {Opt. Express}\ }\textbf {\bibinfo {volume} {8}},\ \bibinfo {pages} {173}
  (\bibinfo {year} {2001})}\BibitemShut {NoStop}%
\bibitem [{\citenamefont {Shen}(2012)}]{shen2012topological}%
  \BibitemOpen
  \bibfield  {author} {\bibinfo {author} {\bibfnamefont {S.-Q.}\ \bibnamefont
  {Shen}},\ }\href@noop {} {\emph {\bibinfo {title} {Topological
  insulators}}},\ Vol.\ \bibinfo {volume} {174}\ (\bibinfo  {publisher}
  {Springer},\ \bibinfo {year} {2012})\BibitemShut {NoStop}%
\bibitem [{\citenamefont {Cheng}\ \emph {et~al.}(2016)\citenamefont {Cheng},
  \citenamefont {Jouvaud}, \citenamefont {Ni}, \citenamefont {Mousavi},
  \citenamefont {Genack},\ and\ \citenamefont {Khanikaev}}]{cheng2016robust}%
  \BibitemOpen
  \bibfield  {author} {\bibinfo {author} {\bibfnamefont {X.}~\bibnamefont
  {Cheng}}, \bibinfo {author} {\bibfnamefont {C.}~\bibnamefont {Jouvaud}},
  \bibinfo {author} {\bibfnamefont {X.}~\bibnamefont {Ni}}, \bibinfo {author}
  {\bibfnamefont {S.~H.}\ \bibnamefont {Mousavi}}, \bibinfo {author}
  {\bibfnamefont {A.~Z.}\ \bibnamefont {Genack}}, \ and\ \bibinfo {author}
  {\bibfnamefont {A.~B.}\ \bibnamefont {Khanikaev}},\ }\href@noop {} {\bibfield
   {journal} {\bibinfo  {journal} {Nat. Mater.}\ }\textbf {\bibinfo {volume}
  {15}},\ \bibinfo {pages} {542} (\bibinfo {year} {2016})}\BibitemShut
  {NoStop}%
\bibitem [{\citenamefont {Chen}\ \emph {et~al.}(2017)\citenamefont {Chen},
  \citenamefont {Zhao}, \citenamefont {Chen},\ and\ \citenamefont
  {Dong}}]{chen2017valley}%
  \BibitemOpen
  \bibfield  {author} {\bibinfo {author} {\bibfnamefont {X.-D.}\ \bibnamefont
  {Chen}}, \bibinfo {author} {\bibfnamefont {F.-L.}\ \bibnamefont {Zhao}},
  \bibinfo {author} {\bibfnamefont {M.}~\bibnamefont {Chen}}, \ and\ \bibinfo
  {author} {\bibfnamefont {J.-W.}\ \bibnamefont {Dong}},\ }\href {\doibase
  10.1103/PhysRevB.96.020202} {\bibfield  {journal} {\bibinfo  {journal} {Phys.
  Rev. B}\ }\textbf {\bibinfo {volume} {96}},\ \bibinfo {pages} {020202}
  (\bibinfo {year} {2017})}\BibitemShut {NoStop}%
\bibitem [{\citenamefont {Oskooi}\ \emph {et~al.}(2010)\citenamefont {Oskooi},
  \citenamefont {Roundy}, \citenamefont {Ibanescu}, \citenamefont {Bermel},
  \citenamefont {Joannopoulos},\ and\ \citenamefont
  {Johnson}}]{oskooi2010meep}%
  \BibitemOpen
  \bibfield  {author} {\bibinfo {author} {\bibfnamefont {A.~F.}\ \bibnamefont
  {Oskooi}}, \bibinfo {author} {\bibfnamefont {D.}~\bibnamefont {Roundy}},
  \bibinfo {author} {\bibfnamefont {M.}~\bibnamefont {Ibanescu}}, \bibinfo
  {author} {\bibfnamefont {P.}~\bibnamefont {Bermel}}, \bibinfo {author}
  {\bibfnamefont {J.~D.}\ \bibnamefont {Joannopoulos}}, \ and\ \bibinfo
  {author} {\bibfnamefont {S.~G.}\ \bibnamefont {Johnson}},\ }\href@noop {}
  {\bibfield  {journal} {\bibinfo  {journal} {Comput. Phys. Commun.}\ }\textbf
  {\bibinfo {volume} {181}},\ \bibinfo {pages} {687} (\bibinfo {year}
  {2010})}\BibitemShut {NoStop}%
\bibitem [{\citenamefont {Hong}\ \emph {et~al.}(1987)\citenamefont {Hong},
  \citenamefont {Ou},\ and\ \citenamefont {Mandel}}]{PhysRevLett.59.2044}%
  \BibitemOpen
  \bibfield  {author} {\bibinfo {author} {\bibfnamefont {C.~K.}\ \bibnamefont
  {Hong}}, \bibinfo {author} {\bibfnamefont {Z.~Y.}\ \bibnamefont {Ou}}, \ and\
  \bibinfo {author} {\bibfnamefont {L.}~\bibnamefont {Mandel}},\ }\href
  {\doibase 10.1103/PhysRevLett.59.2044} {\bibfield  {journal} {\bibinfo
  {journal} {Phys. Rev. Lett.}\ }\textbf {\bibinfo {volume} {59}},\ \bibinfo
  {pages} {2044} (\bibinfo {year} {1987})}\BibitemShut {NoStop}%
\bibitem [{\citenamefont {Ghosh}\ and\ \citenamefont
  {Mandel}(1987)}]{Ghosh1987ObservationON}%
  \BibitemOpen
  \bibfield  {author} {\bibinfo {author} {\bibnamefont {Ghosh}}\ and\ \bibinfo
  {author} {\bibnamefont {Mandel}},\ }\href@noop {} {\bibfield  {journal}
  {\bibinfo  {journal} {Phys. Rev. Lett.}\ }\textbf {\bibinfo {volume} {59}},\
  \bibinfo {pages} {1903} (\bibinfo {year} {1987})}\BibitemShut {NoStop}%
\bibitem [{\citenamefont {Santori}\ \emph {et~al.}(2002)\citenamefont
  {Santori}, \citenamefont {Fattal}, \citenamefont {Vu{\v{c}}kovi{\'c}},
  \citenamefont {Solomon},\ and\ \citenamefont
  {Yamamoto}}]{santori2002indistinguishable}%
  \BibitemOpen
  \bibfield  {author} {\bibinfo {author} {\bibfnamefont {C.}~\bibnamefont
  {Santori}}, \bibinfo {author} {\bibfnamefont {D.}~\bibnamefont {Fattal}},
  \bibinfo {author} {\bibfnamefont {J.}~\bibnamefont {Vu{\v{c}}kovi{\'c}}},
  \bibinfo {author} {\bibfnamefont {G.~S.}\ \bibnamefont {Solomon}}, \ and\
  \bibinfo {author} {\bibfnamefont {Y.}~\bibnamefont {Yamamoto}},\ }\href@noop
  {} {\bibfield  {journal} {\bibinfo  {journal} {Nature}\ }\textbf {\bibinfo
  {volume} {419}},\ \bibinfo {pages} {594} (\bibinfo {year}
  {2002})}\BibitemShut {NoStop}%
\bibitem [{\citenamefont {Wang}\ \emph {et~al.}(2016)\citenamefont {Wang},
  \citenamefont {Chen}, \citenamefont {Li}, \citenamefont {Huang},
  \citenamefont {Liu}, \citenamefont {Chen}, \citenamefont {Luo}, \citenamefont
  {Su}, \citenamefont {Wu}, \citenamefont {Li}, \citenamefont {Lu},
  \citenamefont {Hu}, \citenamefont {Jiang}, \citenamefont {Peng},
  \citenamefont {Li}, \citenamefont {Liu}, \citenamefont {Chen}, \citenamefont
  {Lu},\ and\ \citenamefont {Pan}}]{Wang2016}%
  \BibitemOpen
  \bibfield  {author} {\bibinfo {author} {\bibfnamefont {X.-L.}\ \bibnamefont
  {Wang}}, \bibinfo {author} {\bibfnamefont {L.-K.}\ \bibnamefont {Chen}},
  \bibinfo {author} {\bibfnamefont {W.}~\bibnamefont {Li}}, \bibinfo {author}
  {\bibfnamefont {H.-L.}\ \bibnamefont {Huang}}, \bibinfo {author}
  {\bibfnamefont {C.}~\bibnamefont {Liu}}, \bibinfo {author} {\bibfnamefont
  {C.}~\bibnamefont {Chen}}, \bibinfo {author} {\bibfnamefont {Y.-H.}\
  \bibnamefont {Luo}}, \bibinfo {author} {\bibfnamefont {Z.-E.}\ \bibnamefont
  {Su}}, \bibinfo {author} {\bibfnamefont {D.}~\bibnamefont {Wu}}, \bibinfo
  {author} {\bibfnamefont {Z.-D.}\ \bibnamefont {Li}}, \bibinfo {author}
  {\bibfnamefont {H.}~\bibnamefont {Lu}}, \bibinfo {author} {\bibfnamefont
  {Y.}~\bibnamefont {Hu}}, \bibinfo {author} {\bibfnamefont {X.}~\bibnamefont
  {Jiang}}, \bibinfo {author} {\bibfnamefont {C.-Z.}\ \bibnamefont {Peng}},
  \bibinfo {author} {\bibfnamefont {L.}~\bibnamefont {Li}}, \bibinfo {author}
  {\bibfnamefont {N.-L.}\ \bibnamefont {Liu}}, \bibinfo {author} {\bibfnamefont
  {Y.-A.}\ \bibnamefont {Chen}}, \bibinfo {author} {\bibfnamefont {C.-Y.}\
  \bibnamefont {Lu}}, \ and\ \bibinfo {author} {\bibfnamefont {J.-W.}\
  \bibnamefont {Pan}},\ }\href {\doibase 10.1103/PhysRevLett.117.210502}
  {\bibfield  {journal} {\bibinfo  {journal} {Phys. Rev. Lett.}\ }\textbf
  {\bibinfo {volume} {117}},\ \bibinfo {pages} {210502} (\bibinfo {year}
  {2016})}\BibitemShut {NoStop}%
\bibitem [{\citenamefont {Shen}\ \emph {et~al.}(2017)\citenamefont {Shen},
  \citenamefont {Harris}, \citenamefont {Skirlo}, \citenamefont {Prabhu},
  \citenamefont {Baehr-Jones}, \citenamefont {Hochberg}, \citenamefont {Sun},
  \citenamefont {Zhao}, \citenamefont {Larochelle},\ and\ \citenamefont
  {Englund}}]{Shen2017Deep}%
  \BibitemOpen
  \bibfield  {author} {\bibinfo {author} {\bibfnamefont {Y.}~\bibnamefont
  {Shen}}, \bibinfo {author} {\bibfnamefont {N.~C.}\ \bibnamefont {Harris}},
  \bibinfo {author} {\bibfnamefont {S.}~\bibnamefont {Skirlo}}, \bibinfo
  {author} {\bibfnamefont {M.}~\bibnamefont {Prabhu}}, \bibinfo {author}
  {\bibfnamefont {T.}~\bibnamefont {Baehr-Jones}}, \bibinfo {author}
  {\bibfnamefont {M.}~\bibnamefont {Hochberg}}, \bibinfo {author}
  {\bibfnamefont {X.}~\bibnamefont {Sun}}, \bibinfo {author} {\bibfnamefont
  {S.}~\bibnamefont {Zhao}}, \bibinfo {author} {\bibfnamefont {H.}~\bibnamefont
  {Larochelle}}, \ and\ \bibinfo {author} {\bibfnamefont {D.~a.}\ \bibnamefont
  {Englund}},\ }\href@noop {} {\bibfield  {journal} {\bibinfo  {journal} {Nat.
  Photonics}\ }\textbf {\bibinfo {volume} {11}},\ \bibinfo {pages} {441}
  (\bibinfo {year} {2017})}\BibitemShut {NoStop}%
\bibitem [{\citenamefont {Qiang}\ \emph {et~al.}(2018)\citenamefont {Qiang},
  \citenamefont {Zhou}, \citenamefont {Wang}, \citenamefont {Wilkes},
  \citenamefont {Loke}, \citenamefont {Ogara}, \citenamefont {Kling},
  \citenamefont {Marshall}, \citenamefont {Santagati}, \citenamefont {Ralph},
  \citenamefont {Wang}, \citenamefont {O'Brien}, \citenamefont {Thompson},\
  and\ \citenamefont {Matthews}}]{Qiang2018}%
  \BibitemOpen
  \bibfield  {author} {\bibinfo {author} {\bibfnamefont {X.}~\bibnamefont
  {Qiang}}, \bibinfo {author} {\bibfnamefont {X.}~\bibnamefont {Zhou}},
  \bibinfo {author} {\bibfnamefont {J.}~\bibnamefont {Wang}}, \bibinfo {author}
  {\bibfnamefont {C.~M.}\ \bibnamefont {Wilkes}}, \bibinfo {author}
  {\bibfnamefont {T.}~\bibnamefont {Loke}}, \bibinfo {author} {\bibfnamefont
  {S.}~\bibnamefont {Ogara}}, \bibinfo {author} {\bibfnamefont
  {L.}~\bibnamefont {Kling}}, \bibinfo {author} {\bibfnamefont {G.~D.}\
  \bibnamefont {Marshall}}, \bibinfo {author} {\bibfnamefont {R.}~\bibnamefont
  {Santagati}}, \bibinfo {author} {\bibfnamefont {T.~C.}\ \bibnamefont
  {Ralph}}, \bibinfo {author} {\bibfnamefont {J.~B.}\ \bibnamefont {Wang}},
  \bibinfo {author} {\bibfnamefont {J.~L.}\ \bibnamefont {O'Brien}}, \bibinfo
  {author} {\bibfnamefont {M.~G.}\ \bibnamefont {Thompson}}, \ and\ \bibinfo
  {author} {\bibfnamefont {J.~C.~F.}\ \bibnamefont {Matthews}},\ }\href@noop {}
  {\bibfield  {journal} {\bibinfo  {journal} {Nat. Photonics}\ }\textbf
  {\bibinfo {volume} {12}},\ \bibinfo {pages} {534} (\bibinfo {year}
  {2018})}\BibitemShut {NoStop}%
\bibitem [{\citenamefont {Bocquillon}\ \emph {et~al.}(2013)\citenamefont
  {Bocquillon}, \citenamefont {Freulon}, \citenamefont {Berroir}, \citenamefont
  {Degiovanni}, \citenamefont {Plaçais}, \citenamefont {Cavanna},
  \citenamefont {gan Jin},\ and\ \citenamefont {F{\`e}ve}}]{Bocquillon2013}%
  \BibitemOpen
  \bibfield  {author} {\bibinfo {author} {\bibfnamefont {E.}~\bibnamefont
  {Bocquillon}}, \bibinfo {author} {\bibfnamefont {V.}~\bibnamefont {Freulon}},
  \bibinfo {author} {\bibfnamefont {J.-M.}\ \bibnamefont {Berroir}}, \bibinfo
  {author} {\bibfnamefont {P.}~\bibnamefont {Degiovanni}}, \bibinfo {author}
  {\bibfnamefont {B.}~\bibnamefont {Plaçais}}, \bibinfo {author}
  {\bibfnamefont {A.}~\bibnamefont {Cavanna}}, \bibinfo {author} {\bibfnamefont
  {Y.}~\bibnamefont {gan Jin}}, \ and\ \bibinfo {author} {\bibfnamefont
  {G.}~\bibnamefont {F{\`e}ve}},\ }\href@noop {} {\bibfield  {journal}
  {\bibinfo  {journal} {Science}\ }\textbf {\bibinfo {volume} {339}},\ \bibinfo
  {pages} {1054} (\bibinfo {year} {2013})}\BibitemShut {NoStop}%
\bibitem [{\citenamefont {Fakonas}\ \emph {et~al.}(2014)\citenamefont
  {Fakonas}, \citenamefont {Lee}, \citenamefont {Kelaita},\ and\ \citenamefont
  {Atwater}}]{Fakonas2014Twoplasmon}%
  \BibitemOpen
  \bibfield  {author} {\bibinfo {author} {\bibfnamefont {J.~S.}\ \bibnamefont
  {Fakonas}}, \bibinfo {author} {\bibfnamefont {H.}~\bibnamefont {Lee}},
  \bibinfo {author} {\bibfnamefont {Y.~A.}\ \bibnamefont {Kelaita}}, \ and\
  \bibinfo {author} {\bibfnamefont {H.~A.}\ \bibnamefont {Atwater}},\
  }\href@noop {} {\bibfield  {journal} {\bibinfo  {journal} {Nat. Photonics}\
  }\textbf {\bibinfo {volume} {8}},\ \bibinfo {pages} {317} (\bibinfo {year}
  {2014})}\BibitemShut {NoStop}%
\bibitem [{\citenamefont {Chen}\ \emph {et~al.}(2018)\citenamefont {Chen},
  \citenamefont {Lee}, \citenamefont {Lu}, \citenamefont {Liu}, \citenamefont
  {Wu}, \citenamefont {Feng}, \citenamefont {Li}, \citenamefont {Rockstuhl},
  \citenamefont {Guo}, \citenamefont {Guo}, \citenamefont {Tame},\ and\
  \citenamefont {Ren}}]{Chen2018QPN}%
  \BibitemOpen
  \bibfield  {author} {\bibinfo {author} {\bibfnamefont {Y.}~\bibnamefont
  {Chen}}, \bibinfo {author} {\bibfnamefont {C.}~\bibnamefont {Lee}}, \bibinfo
  {author} {\bibfnamefont {L.}~\bibnamefont {Lu}}, \bibinfo {author}
  {\bibfnamefont {D.}~\bibnamefont {Liu}}, \bibinfo {author} {\bibfnamefont
  {Y.}~\bibnamefont {Wu}}, \bibinfo {author} {\bibfnamefont {L.-T.}\
  \bibnamefont {Feng}}, \bibinfo {author} {\bibfnamefont {M.}~\bibnamefont
  {Li}}, \bibinfo {author} {\bibfnamefont {C.}~\bibnamefont {Rockstuhl}},
  \bibinfo {author} {\bibfnamefont {G.}~\bibnamefont {Guo}}, \bibinfo {author}
  {\bibfnamefont {G.}~\bibnamefont {Guo}}, \bibinfo {author} {\bibfnamefont
  {M.}~\bibnamefont {Tame}}, \ and\ \bibinfo {author} {\bibfnamefont {X.-F.}\
  \bibnamefont {Ren}},\ }\href@noop {} {\bibfield  {journal} {\bibinfo
  {journal} {Optica}\ }\textbf {\bibinfo {volume} {5}},\ \bibinfo {pages}
  {1229} (\bibinfo {year} {2018})}\BibitemShut {NoStop}%
\bibitem [{\citenamefont {Toyoda}\ \emph {et~al.}(2015)\citenamefont {Toyoda},
  \citenamefont {Hiji}, \citenamefont {Noguchi},\ and\ \citenamefont
  {Urabe}}]{Toyoda2015}%
  \BibitemOpen
  \bibfield  {author} {\bibinfo {author} {\bibfnamefont {K.}~\bibnamefont
  {Toyoda}}, \bibinfo {author} {\bibfnamefont {R.}~\bibnamefont {Hiji}},
  \bibinfo {author} {\bibfnamefont {A.}~\bibnamefont {Noguchi}}, \ and\
  \bibinfo {author} {\bibfnamefont {S.}~\bibnamefont {Urabe}},\ }\href@noop {}
  {\bibfield  {journal} {\bibinfo  {journal} {Nature}\ }\textbf {\bibinfo
  {volume} {527}},\ \bibinfo {pages} {74} (\bibinfo {year} {2015})}\BibitemShut
  {NoStop}%
\bibitem [{\citenamefont {Lopes}\ \emph {et~al.}(2015)\citenamefont {Lopes},
  \citenamefont {Imanaliev}, \citenamefont {Aspect}, \citenamefont {Cheneau},
  \citenamefont {Boiron},\ and\ \citenamefont {Westbrook}}]{Lopes2015Atomic}%
  \BibitemOpen
  \bibfield  {author} {\bibinfo {author} {\bibfnamefont {R.}~\bibnamefont
  {Lopes}}, \bibinfo {author} {\bibfnamefont {A.}~\bibnamefont {Imanaliev}},
  \bibinfo {author} {\bibfnamefont {A.}~\bibnamefont {Aspect}}, \bibinfo
  {author} {\bibfnamefont {M.}~\bibnamefont {Cheneau}}, \bibinfo {author}
  {\bibfnamefont {D.}~\bibnamefont {Boiron}}, \ and\ \bibinfo {author}
  {\bibfnamefont {C.~I.}\ \bibnamefont {Westbrook}},\ }\href@noop {} {\bibfield
   {journal} {\bibinfo  {journal} {Nature}\ }\textbf {\bibinfo {volume}
  {520}},\ \bibinfo {pages} {66} (\bibinfo {year} {2015})}\BibitemShut
  {NoStop}%
\bibitem [{\citenamefont {Vest}\ \emph {et~al.}(2017)\citenamefont {Vest},
  \citenamefont {Dheur}, \citenamefont {Devaux}, \citenamefont {Baron},
  \citenamefont {Rousseau}, \citenamefont {Hugonin}, \citenamefont {Greffet},
  \citenamefont {Messin},\ and\ \citenamefont {Marquier}}]{Vest1373}%
  \BibitemOpen
  \bibfield  {author} {\bibinfo {author} {\bibfnamefont {B.}~\bibnamefont
  {Vest}}, \bibinfo {author} {\bibfnamefont {M.-C.}\ \bibnamefont {Dheur}},
  \bibinfo {author} {\bibfnamefont {{\'E}.}~\bibnamefont {Devaux}}, \bibinfo
  {author} {\bibfnamefont {A.}~\bibnamefont {Baron}}, \bibinfo {author}
  {\bibfnamefont {E.}~\bibnamefont {Rousseau}}, \bibinfo {author}
  {\bibfnamefont {J.-P.}\ \bibnamefont {Hugonin}}, \bibinfo {author}
  {\bibfnamefont {J.-J.}\ \bibnamefont {Greffet}}, \bibinfo {author}
  {\bibfnamefont {G.}~\bibnamefont {Messin}}, \ and\ \bibinfo {author}
  {\bibfnamefont {F.}~\bibnamefont {Marquier}},\ }\href {\doibase
  10.1126/science.aam9353} {\bibfield  {journal} {\bibinfo  {journal}
  {Science}\ }\textbf {\bibinfo {volume} {356}},\ \bibinfo {pages} {1373}
  (\bibinfo {year} {2017})}\BibitemShut {NoStop}%
\end{thebibliography}%

\end{document}